\documentclass[reprint, magnetization amssymb, superscriptaddress, longbibliography ]{revtex4-2}
\usepackage{graphicx}
\usepackage[utf8]{inputenc}
\usepackage{lineno}
\usepackage{appendix}
\usepackage{amsmath}

\begin{document}
\author{M.\ Bia\l{}ek}
\email{marcin.bialek@unipress.waw.pl}
\address{Institute of High Pressure Physics, Polish Academy of Sciences, Warszawa, Poland}
\author{Y. Todorov}
\address{Laboratoire de Physique de l’Ecole Normale Supérieure, ENS, Université PSL, CNRS, Sorbonne Université, Université Paris Cité, Paris, France}
\author{K.\ Stelmaszczyk}
\address{Institute of High Pressure Physics, Polish Academy of Sciences, Warszawa, Poland}
\author{D.\ Szwagierczak}\author{B.\ Synkiewicz-Musialska}\author{J.\ Kulawik}
\address{
Łukasiewicz Research Network–Institute of Microelectronics and Photonics, Kraków Division, Kraków, Poland}
\author{N.\ Pałka}
\address{Institute of Optoelectronics, Military University of Technology, Warszawa, Poland}
\author{M.\ Potemski}
\address{LNCMI-EMFL, CNRS UPR3228, Univ. Grenoble Alpes, Univ. Toulouse, Univ. Toulouse 3, INSA-T, Grenoble and Toulouse, France}
\address{CENTERA, CEZAMAT, Warsaw University of Technology, 02-822 Warszawa, Poland}
\address{Institute of Experimental Physics, Faculty of Physics, University of Warsaw, Warszawa, Poland}
\author{W.\ Knap}
\address{Institute of High Pressure Physics, Polish Academy of Sciences, Warszawa, Poland}
\address{CENTERA, CEZAMAT, Warsaw University of Technology, 02-822 Warszawa, Poland}

\title{Hybridization of terahertz phonons and magnons in disparate and spatially-separated material specimens}
\date{\today}

\begin{abstract}
The interaction between light and matter in condensed matter excitations and
electromagnetic resonators serves as a rich playground for fundamental research
and lies at the core of photonic and quantum technologies. Herein, we present
comprehensive experimental and theoretical studies of the photon-mediated
hybridization of magnons and phonons in the terahertz (THz) range. We demonstrate
the intriguing concept of composite states formed by distinct electric and magnetic
quasiparticles strongly coupled to the same optical cavity modes. Specifically, we
explore magnons excited in a slab of an antiferromagnetic crystal and phonons
excited in a distinct specimen of an insulating material.
The crystal slabs form an optical cavity with Fabry-P\'erot oscillations in the THz
range. We demonstrate hybridized phonon-magnon polariton modes and their
tunability by adjusting the distance between the slabs, showing that hybridization
persists even at separations up to several millimeters. The experimental results are
interpreted using both classical and quantum electrodynamical models. The quantum
description allows us to quantify the degree of hybridization that is linked to a
topological behavior of the electric field phasor, in agreement with the classical
electrodynamics expectations. Importantly, the presented results refer to temperature
conditions and cavities of millimeter size, paving the way for engineering realistic,
frequency-tunable THz devices through the hybridization of electric (phononics) and
magnetic (spintronics) elementary excitations of matter.
\end{abstract}
\maketitle

In the regime of strong coupling between matter excitations and electromagnetic cavities the energies of light and matter excitations are periodically exchanged, resulting in mixed light-matter coupled quasiparticles known as polaritons \cite{Khitrova06, Dovzhenko18}.
Light-matter coupling was first reported with single atoms \cite{Rempe87}, and later with single excitons bound in quantum dots \cite{Khitrova06, Kasprzak06}.
The light-matter coupling strength of $N$ oscillators placed in a cavity is enhanced by a factor of $\sqrt{N}$, a phenomenon known as Dicke cooperativity \cite{Raizen89, Colombe07, Delteil_prl2012, Torma14, Basov16, Bayer17, Li18, Yahiaoui22}. It allows the achievement of observable couplings of magnetization oscillations despite a tiny dipolar moment of a single spin.
The strong coupling of microwave cavity modes and magnons was observed as early as 1962 \cite{Roberts62}, and since the 2010s it has been intensively studied \cite{Schuster10, Abe11, Huebl13, Zhang14, Tabuchi14, Tabuchi15, Zhang15, Zhang16, Li19, Everts20, Potts20, Lachance-Quirion20, Li20JAP, Bhoi21}, because of the prospect of using magnon-polaritons in quantum devices \cite{Awschalom07, Kockum19, Roux20, Yuan22}. 

Strong light-matter coupling can form hybridized states of two distant resonators exchanging energy with the same electromagnetic cavity mode.
Such light-bound composite states were shown for different electric dipole resonators \cite{Spethmann16, Sciesiek20, Yahiaoui22}. The coupling of two ferromagnetic magnons mediated by a cavity mode was recently achieved using superconducting circuits \cite{Xu19, Li22} and nanostripline antennas \cite{Hanchen22}, and was discussed theoretically \cite{Xu19, Harvey-Collard22, Nair22, Yang22}. 
We recently showed that magnons in distant antiferromagnets can be coupled using Fabry-P\'erot cavity modes \cite{Bialek23}.

An even more intriguing concept is the possibility of achieving composite states of two different optically active quasiparticles, such as phonons, magnons, collective excitations such as plasmons, or Landau level excitations, many of which have energies in the THz range.
Such couplings of magnons are commonly observed with other excitations in the same material, such as acoustic phonons \cite{Berk19, Li21}, optical phonons in a multiferroic \cite{Khan20} and in layered magnets \cite{Li20PRL, Liu21, Vaclakova21}, excitons in a layered semiconductor \cite{Diederich22}, and electron paramagnetic resonance \cite{Li18}. 
However, photon-mediated coupling of distinct excitations in different materials, was only shown in the weak regime, i.e.\ without frequency matching of phonons and magnons, in a heterostructure of an antiferromagnet and lithium niobate (LiNbO$_3$) \cite{Sivarajah19}.

Here, we demonstrate frequency-matched coupling of antiferromagnetic magnons and optical phonons hosted exclusively in two different and distant samples. This coupling is mediated by cavity modes formed by the sample slabs themselves.
\begin{figure}
\begin{center}
\includegraphics[width=0.9\linewidth]{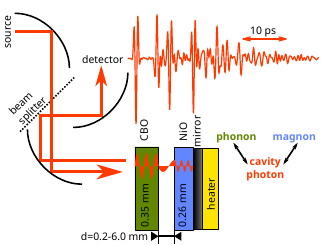}
\caption{{\label{setup}
The measured system comprised NiO crystal placed on a copper mirror and a slab of CBO. We measured reflection as a function of the gap $d$ between the slabs and the temperature $T$ of the NiO crystal. In the top-right corner of the figure, we show a sample time trace.}}
\end{center}
\end{figure}
Experiments were performed close to room temperature and without external fields. These two sample slabs were placed next to each other at a well-controlled gap, and the slabs themselves formed the cavity, that is, the cavity fields penetrated both materials (Fig.\ \ref{setup}). 
We show phonon-magnon-polariton modes, under frequency matching of magnon, phonon, and cavity mode.
Such hybrid modes have magnetic and electric dipole moments originating from their constituents.
We show that our experimental data can be interpreted with a model based on classical electrodynamics, and a microscopic model. The magnon-photon coupling remains observable when the crystal separation reaches the millimeter range. The separate magnon and phonon modes are coupled due to the strong coupling of each resonance to the same mode of a Fabry-P\'erot cavity. 
This research demonstrates hybrid resonators that could be used in communication, quantum technologies, or for the detection of hypothetical axion particles \cite{Marsh19, Mitridate20}, using materials hosting topological magnons \cite{Wang11, Karaki22} with the advantage of large resonator volumes and tunable detection frequency \cite{Marsh22}.

We used a single crystal of nickel oxide (NiO) 111 cut of thickness $h_m=0.26$ mm and lateral diameter of 5~mm. The antiferromagnetic resonance in this antiferromagnetic material, the $\alpha$-mode \cite{Rezende19}, has a frequency close to 1.0 THz at room temperature. It decreases with rising temperature, reaching 0.7 THz at about 450 K. We chose ceramics of Cu$B_2$O$_4$ (CBO) \cite{Szwagierczak21}, as a material that hosts a relatively narrow phonon mode at about 0.92 THz, which can be crossed with the magnon in NiO.
The oscillator strength of the phonon in CBO is weak for an electric dipole excitation, while the oscillator strength of the magnon in NiO is quite strong for a magnetic dipole. This allows for comparable light-matter coupling strengths in both cases, so the electric dipole excitation does not completely dominate the magnetic dipole excitation.
The NiO crystal was fixed on a copper mirror, the temperature of which was measured with a K-type thermocouple. Temperature was stabilized with a software proportional–integral–derivative controller setting electric current applied to a stack of Peltier elements beneath the copper mirror. The NiO crystal and CBO ceramics were placed on independent kinematic mounts, allowing for precise parallelism control. The size $d$ of the gap between the samples was controlled with a motorized stage that supports the NiO part of the cavity. Thermal transfer due to convection was small for $d\gtrsim 0.2$~mm, and the phonon frequency does not depend on temperature. The experimental setup was purged with dry air to minimize the effect of atmospheric water absorption with a few strong lines around 1.0 THz.

We used the THz time-domain spectrometer system (Toptica). Optical pulses were guided in single-mode zero-dispersion fibers, and a biased source antenna was emitting pulses of electromagnetic radiation with spectrum power centered around 1.0 THz. The THz beam was propagated from the source using a quasi-optical setup with parabolic mirrors. We used a beam splitter made of Kapton tape, which directs pulses reflected from the sample to the detector antenna. We used a 200 ps delay line, which gives a 5 GHz resolution. In the presented spectra, we extended these time-domain data to 1000 ps with zeros, interpolating the frequency-domain resolution to 1 GHz. Spectra were collected as a function of gap $d$ between the crystals for a fixed NiO temperature $T$. Then, the temperature was changed with a step of 1.0~K and spectra were collected again as a function of the gap $d$. We Fourier-transform measured time traces of electric field, which yields complex spectra. This report shows the standard magnitude of obtained spectra, clearly showing strong coupling and magnons, phonons, and cavity modes.
\begin{figure}
\includegraphics[width=\linewidth]{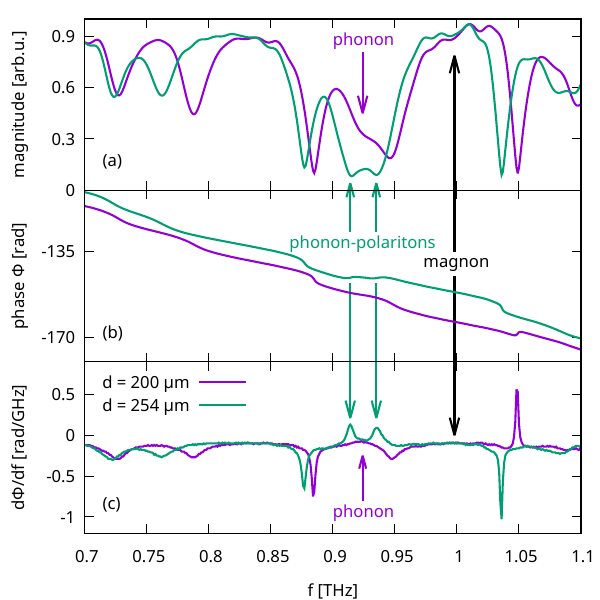}
\caption{\label{dPdf}
 Comparison of the (a) magnitude spectrum with its (b) phase and (c) phase differential to frequency. The latter shows typically, much narrower features around resonances. Also, the phase jump can be positive or negative, whereas optical and matter resonances always result in a minimum amplitude of reflection spectra.
}
\end{figure}

\begin{figure}
\includegraphics[width=\linewidth]{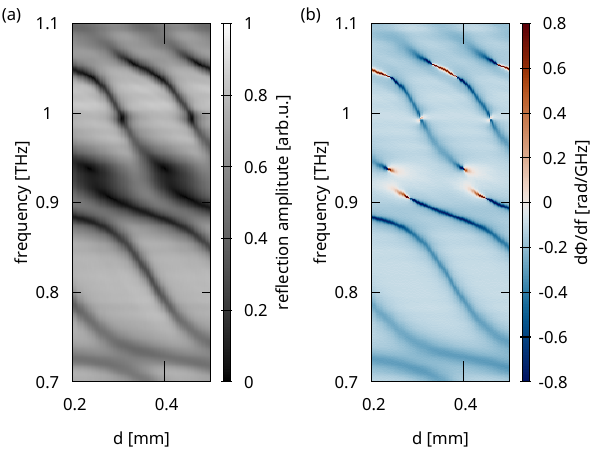}
\caption{\label{mag-pha-comparison}
 Reflection spectra: (a) amplitude and (b) phase differential to frequency. Spectra are plotted as a function of the gap $d$ between the slabs. 
}
\end{figure}

However, often omitted, phase carries more information of better quality than magnitude. This is mostly because spectral features in phase are narrower than in magnitude, and phase jumps at resonances can have positive or negative signs. In our convention, the reflection phase $\Phi$ drops almost linearly with rising frequency, as expected in the case of electromagnetic waves propagating in a non-dispersive infinite medium. We remove this constant trend by calculating phase differential to frequency $d\Phi/df$, in the unit of rad/GHz. This treatment of phase spectra reveals rapid phase changes characterizing resonances, both cavity modes and optically active transitions in the solid state such as phonons or magnons. Such phase data do not require any further normalization. In our spectra, we observe that the sign of optical resonances is typically negative, that is, in the direction of linear change of phase with increasing frequency. On the other hand, the signs of NiO magnon and CBO phonon phase jumps are always positive in our convention (white and red in \ref{mag-pha-comparison}). Such pure matter modes are barely visible due to either, the low amplitude of the magnon, or the large width of the phonon. However, they become much more pronounced when they couple with cavity modes. Moreover, we observed that the sign of polariton modes is also always positive in the strong-coupling regime. We observed, that the sign of phase jump is an indicator of the strong coupling, and only cavity modes with positive phase produce show strong coupling, as visible in Fig.\ \ref{mag-pha-comparison}b, where the magnon just below 1.0 THz shows only weak coupling with FP modes and cavity sign is negative (blue), while the phonon forms polaritons modes with FP modes and its sign is positive (red).

Mode frequencies of the FP cavity are lowering, with the rising gap $d$ (Fig.\ \ref{mag-pha-comparison}). When they cross with the CBO phonon mode at around 0.92 THz, they form two phonon-polariton modes, about 20 GHz apart. In Fig.\ \ref{mag-pha-comparison}b phase jumps of these polariton modes are characterized by positive jumps (red color). By our convention, cavity modes have typically negative signs represented with blue colors (Fig.\ \ref{mag-pha-comparison}). However, some strong cavity modes may have positive signs, like the one at around 1.05 THz. Magnon mode in NiO is only slightly visible near 1.0 THz because the results were obtained at 300 K. In this case, the interaction of the magnon with FP cavity modes is weak, resulting in only small splittings, comparable with polariton widths.

\begin{figure}
\includegraphics[width=\linewidth]{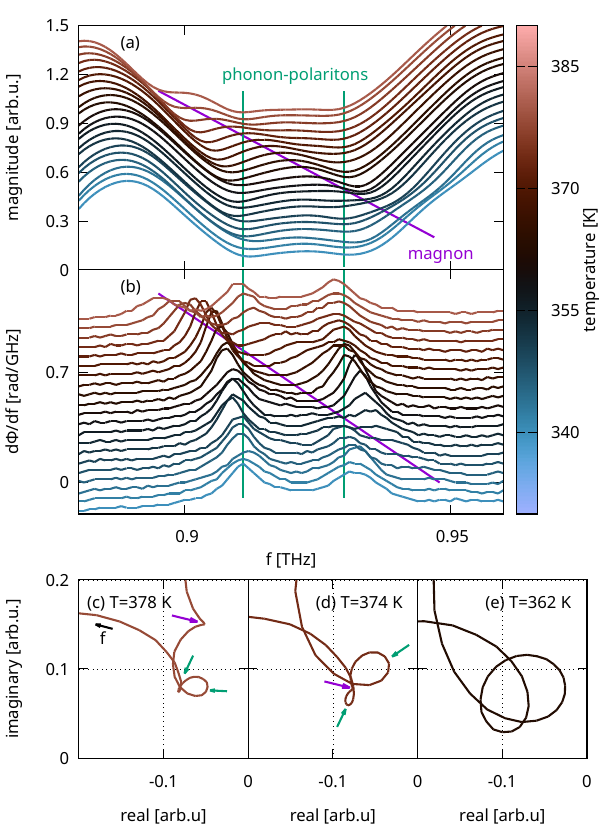}
\caption{\label{mag-phon-cut}
 Reflection spectra showing the influence of NiO magnon on phonon-polaritons. Magnon frequency sweep was achieved by changing NiO crystal temperature. Magnitude (a) and phase (b) of reflection spectra. Green lines guide the eye showing phonon-polaritons, and a violet line indicates the magnon mode. 
 Phasor plots (c,d) in the range of 985-940 GHz for different temperatures, show a double loop indicating phonon polaritons (green arrows), and a feature indicating magnon (violet arrow). Phasor plot (e) showing phonon-magnon polaritons. 
}
\end{figure}

The magnon frequency in NiO was tuned in the range of about 0.7--1.0 THz, using its dependence on crystal temperature in the range of about 300--450 K. When the magnon frequency passes the frequency of the phonon in CBO (0.92 THz), these two matter modes would cross without any interaction, as visible in Fig. \ref{pha-T}g ($d=305$ $\mu$m). However, when the gap between the slabs is such that one of the cavity modes has a frequency close to the CBO phonon, we observe phonon-magnon-polariton modes as a function of NiO temperature. At a particular distance of $d=260$ $\mu$m (Fig. \ref{pha-T}d), it is visible that the magnon mode reaches the phonon frequency at about $T=360$ K. The magnon interacts with the phonon-polaritons, making them narrower and more separated. The narrowing is because magnons have a line width much smaller than that of phonons, thus, owing to hybridization, the resulting phonon-magnon-polaritons are narrower than pure phonon-polaritons. Phonon-magnon polaritons are also slightly more split than pure phonon-polaritons because of magnons' small additional oscillator strength. It is also shown in Fig.\ \ref{mag-phon-cut}ab, where green lines were measured for a magnon-phonon matching condition ($T=361$ K), while the violet curve shows pure phonon-polaritons with magnon mode around 0.96 THz. A small change of the gap $d$ between the slabs compensates for the drift of dielectric constants of both materials with temperature. In Fig.\ \ref{mag-phon-cut}cd, we show pure magnon-polariton modes around 0.89 THz, and pure phonon mode around 0.92 THz.

Phase data can also be analyzed with amplitude data, using phasor plots of the imaginary part of the electric field as a function of its real part, as in Fig.\ \ref{mag-phon-cut}ce. Normally, the signal circles the origin with rising frequency, which reflects phase progression. Magnon and phonon modes are very weakly distinguishable as phase progression slowdowns and signal amplitude drops related to absorption. In cavity modes, the circle progresses faster and closer to the origin.
In some cavity modes, the loop becomes so tight that it no longer circles the origin, resulting in a reversed-phase progression, such as the modes around 1.05 THz in Fig.\ \ref{mag-pha-comparison}b (more details in the Appendix \ref{Phase-sign}). 
The phasor signature of the strong coupling is two intertwined loops \cite{Thomas20}, and neither of them encircles the origin point. Thus, the phase progression reverses twice during this signal evolution (green arrows indicate phonon-polaritons in Fig.\ \ref{mag-phon-cut}c). 
We also demonstrate in detail the formation of phonon-polaritons phasor signature in Appendix \ref{Phase-sign}, when the magnon mode was spectrally distant from the phonon.
In Fig.\ \ref{mag-phon-cut}c-e, we show the influence of magnon mode (features indicated by violet arrows) on phonon polaritons (loops indicated by green arrows) 
The loops indicating the strong coupling are much smaller for phonon-polaritons (green arrows in Fig.\ \ref{mag-phon-cut}cd) than for phonon-magnon-polaritons (Fig.\ \ref{mag-phon-cut}e). This is related to the modes narrowing and deepening, caused by the hybridization with magnon. In contrast to Fig.\ \ref{mag-phon-cut}cd, in Fig.\ \ref{mag-phon-cut}e, the magnon signature cannot be any more distinguished otherwise than the increase of the loops, which indicates the hybridization of phonons and magnons.

Having demonstrated the strong coupling between magnons, phonons, and Fabry-P\'erot resonances experimentally, we now quantitatively investigate the mixing between them. This mixing is best described within the framework of a quantum model which treats the system as coupled quantum harmonic oscillators. This model has been detailed in the Appendix \ref{AnnexQM}. Starting from a general quantum electrodynamical description the total Hamiltonian of the tripartite system is derived to be:
\begin{multline}
\label{Hamilatonian}
    \mathcal{H}=\hbar\omega_m(a^\dag a+1/2)+\hbar\omega_{LO}d^\dag d+\hbar\omega_\alpha c^\dag c\\+i\hbar\Omega_p(a-a^\dag)(d+d^\dag)+i\hbar\Omega_m(c-c^\dag)(a+a^\dag),
\end{multline}
This Hamiltonian was derived by considering the coupling of the electric field of the FP resonance with the phonon polarization. In contrast, the magnetic field is coupled to the magnon magnetization field through a Zeeman interaction.
Here $\omega_m$ is 
the frequency of the Fabry-P\'erot resonance,
$\omega_{LO}$ is the LO optical phonon frequency of the CBO slab, and $\omega_{\alpha}$ is the $\alpha-$magnon frequency. The operator $a$ is the bosonic destruction operator for the photon mode and $d$ and $c$ are collective boson operators describing the phonon and magnon excitations in the two slabs. In the above equation, we have introduced the phonon-photon coupling rate $\Omega_p$ and the magnon-photon coupling rate, $\Omega_m$. As shown in Annex \ref{AnnexQM} they are provided by:
\begin{equation}
\label{Omega_p}
    \hbar\Omega_p=\frac{1}{2}\hbar\eta_{p}\sqrt{\frac{(\omega_{LO}^2-\omega_{TO}^2)\omega_m}{\omega_{LO}}}
\end{equation}
and
\begin{equation}
\label{Omega_m}
    \hbar\Omega_m=\frac{1}{2}\left(\frac{H_{Ax}}{H_E}\right)^{\frac{1}{4}}\eta_m g\mu_B\sqrt{\frac{\hbar\omega_m\mu_0}{2}\left(\frac{N}{V_{NiO}}\right)}.
\end{equation}

We recall the transverse and longitudinal optical photon frequencies are linked by the relation $\omega_{TO}=\omega_{LO}\sqrt{\frac{\epsilon_{\infty}}{\epsilon_{st}}}$, where the dielectric constants have been determined experimentally as described further.


The collective nature of the magnon-light coupling is obvious from the expression of the coupling constant, Eq.(\ref{Omega_m}), which is proportional to the square of the spin density $N/V_{NiO}$.
In NiO, $\mu_0H_{Ax}\approx0.5$~T is the easyplane anisotropy field at room temperature, $\mu_0H_E=698$~T is the exchange field \cite{Rezende19}, and spin density $N/V_{NiO}=5.489\times10^{28}$~m$^{-3}$. \cite{Roth58, Massarotti91} 
By diagonalizing the Hamiltonian (1) through the Hopfield-Bogoliubov procedure we obtain the dispersion relation for the polariton frequencies and closed expressions for the mixing coefficients (Appendix \ref{AnnexQM}). To match the coupled mode frequencies with the experimental ones, we have considered the overlap factors $\eta_p, \eta_m$ as the fitting parameters.
\begin{figure}
\includegraphics[width=\linewidth]{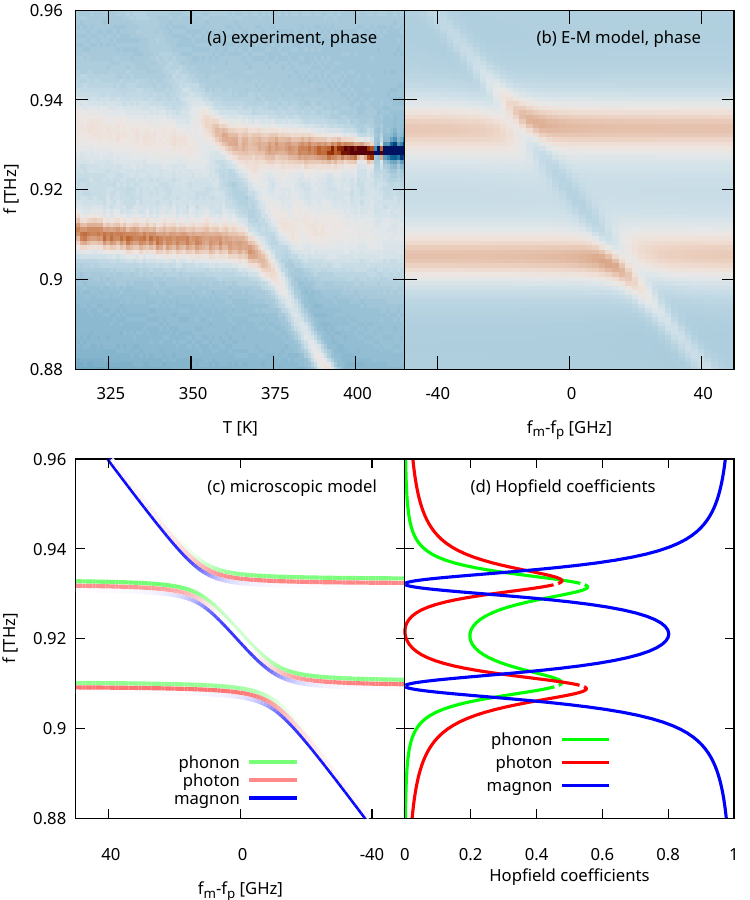}
\caption{\label{quantum}
 Measured phase differential to frequency of reflection spectra plotted as a function of NiO temperature $T$, for a matching mode frequency (a). Predictions of the electromagnetic model (b). Predictions of interacting mode frequencies, and magnon, phonon, and photon contents using the microscopic model (Eq.\ \ref{Hamilatonian}) (c). Hopfield coefficients for photon, magnon, and phonon (d). 
}
\end{figure}
In Fig.\ \ref{quantum}c, we use effective coefficients $\eta_p=.4$, and $\eta_m=0.6$, which better reproduce the frequencies observed in the experimental spectra (Fig.\ \ref{quantum}a). Calculations based on the microscopic model show that the middle mode, under the frequency matching condition $f_c=f_m=f_p$, is composed in about 80\% of magnon and in 20\% of phonon, while it has a less than 0. 3\% photon content. Unequal contents of the phonon and the magnon in this mode result from their unequal coupling strengths to a cavity mode. This property does not change much when tuning the magnon frequency within about 50\% of the phonon-polariton gap. However, in the experiment (Fig.\ \ref{quantum}a) this mode has a small positive amplitude, which is not predicted using the microscopic model. At the same condition of frequency matching, two other modes contain approximately equally 45\% of phonon and photon, while only about 10\% of magnon. Modes of larger amplitude and with more equal contents of phonon and magnon can be observed when the magnon mode has a frequency close to either upper or lower phonon-polaritons, that is, in Fig.\ \ref{quantum}, $|f_m-f_p|\approx 12$ GHz. Then, two phonon-magnon polariton modes with a smaller splitting are formed, which both contain about 40\% phonon and magnon and only 20\% photons. Due to a smaller splitting, they are strongly tunable with magnon frequency in terms of magnon and phonon content.


\begin{figure*}
\includegraphics[width=\linewidth]{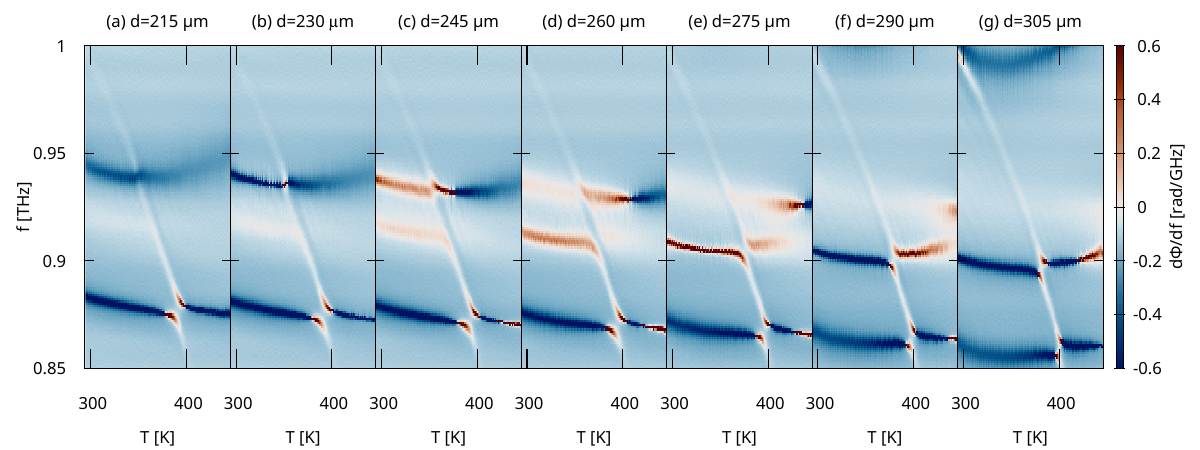}
\includegraphics[width=\linewidth]{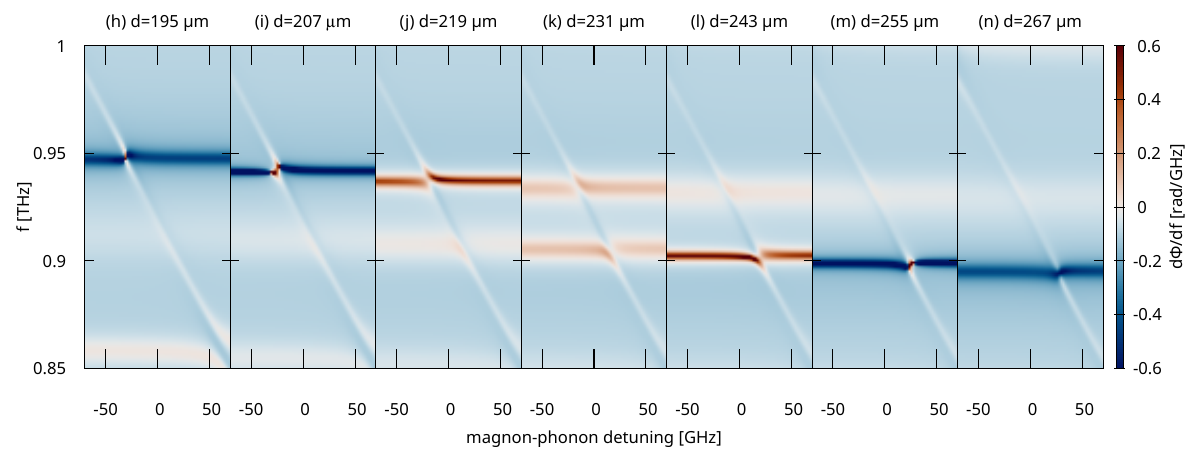}
\caption{\label{pha-T}
 Measured phase differential to the frequency of reflection spectra plotted as a function of NiO temperature $T$, for 7 different gaps $d$ between the slabs (a-g). 
 Predicted spectra of reflection phase differential to frequency, as a function of the detuning of the NiO magnon from the frequency of the CBO phonon, for 7 different gaps $d$ between the slabs (h-n).
}
\end{figure*}
The reflection spectra can be modeled using the expressions for the dielectric and magnetic susceptibilities of the layers, which were determined using transmission measurements of each of the layers (Appendix \ref{dielectric_characterisation}). This approach allows a direct comparison of the amplitude and phase-resolved measurement for the electric field as obtained by the THz-TDs experiment. The compatibility of this approach and the quantum description are commented in the Appendix \ref{AnnexQM}. 
We analytically calculated reflection from a series of parallel-plane slabs using a characteristic matrix method for isotropic media \cite{BORN}, details of which are explained in Ref.\ \onlinecite{Bialek23}. We assumed a Lorentzian shape of both magnon and phonon modes, and otherwise constant background permittivity in the range of interest (0.5-1.5 THz). The CBO sample is ceramics, so it is right to assume that it is isotropic. In the case of the NiO crystal, $\mu$ does not contain any off-diagonal terms, therefore, we can consider it isotropic in our model. Details of assumed permittivity and permeability are shown in Appendix A.

\begin{figure}
\includegraphics[width=\linewidth]{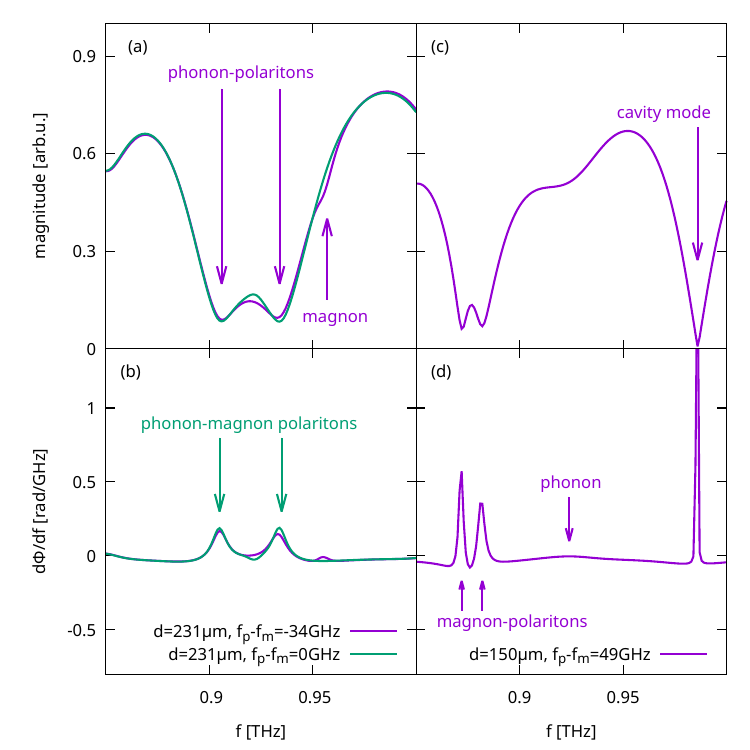}
\caption{\label{mag-phon-cut-theo}
The predicted reflection spectra are calculated using model parameters that roughly reproduce the experimental data presented in Fig.\ \ref{mag-phon-cut}.
Magnitude (a) and phase (b) spectra show phonon-polaritons (violet) for the magnon frequency away from the phonon and phonon-magnon-polaritons (green) for zero detuning.
Magnitude (c) and phase (d) spectra, showing magnon-polaritons around 0.88 THz.
}
\end{figure}

The predicted reflection phase spectra are shown in Fig.\ \ref{pha-T}(h-n) as a function of magnon-phonon frequency detuning ($f_p-f_m$) for a few different gaps between the modeled slabs. We used a simplified model where cavities are independent of the detuning (temperature in the experiment). Qualitatively, this model predicts a similar effect as measured as a function of temperature (Fig.\ \ref{pha-T}(a-g)). The model predicts that the gap, at which cavity mode matches phonon frequency, is different from measured values by 29 $\mu$m, which might stem from inaccuracies of thickness measurement, gap measurement, and dielectric constant determination. The model correctly predicts a change in the phase jump sign of polaritons when they interact strongly with either phonons or magnons. No magnon-phonon coupling is predicted when none of the cavity modes matches the phonon mode (Fig.\ \ref{pha-T}gn). In Fig.\ \ref{mag-phon-cut-theo}, we present some predicted spectra, which are chosen for parameters approximately reproducing the experimental spectra presented in Fig.\ \ref{mag-phon-cut}de and Fig.\ \ref{dPdf}. Panels a and b of Fig.\ \ref{mag-phon-cut-theo} show the influence of the magnon on the phonon-polariton modes, with spectra similar to those in Fig.\ \ref{dPdf}. The influence of magnon on phonon-polaritons is visible as a narrowing of the modes and a small increase in the splitting. Panels c and d of Fig.\ \ref{mag-phon-cut-theo} show magnon-polaritons at about 0.88 THz, away from the phonon mode at 0.92 THz. Here phonon does not interact, because there is no cavity mode at a close frequency.

In summary, we presented systematic experimental studies on the interaction of two distinct polariton systems as a function of temperature and separation distance. The cooperative interaction of magnons in the slab of nickel oxide (NiO) and phonons in the distant slab of CuB$_2$O$_4$ (CBO) is mediated by a cavity mode, which leads to the formation of phonon-magnon polaritons states. The experiments were performed at room temperatures, and the separation between the crystals was in the millimeter range, i.e. showing the potential of this coupling scheme for applications. We used the electromagnetic model, which describes the observed features with realistic assumptions for the parameter values.
Cavity-mediated coupling enables control of antiferromagnetic polaritons by modulating the resonator surroundings. It shows the possibility of designing hybrid resonators operating in the THz range (few meV) that share properties of their components, such as width, amplitude, and dependence on external parameters. 

More generally, our results present an even more intriguing concept of composite states made of optically active quasiparticles of distinct nature, which exchange the same cavity photon in the strong coupling regime. These excitations could be plasmons in quantum confined electron gases \cite{Todorov_PhysRevB125409_2015, Ghosh21}, or intersubband cavity polaritons that can reach the ultra-strong coupling regime \cite{Todorov_PRL2010}. A very appealing perspective is realizing such hybridization with THz meta-atom resonators that can provide strong confinement for both the electric and magnetic field \cite{Jeannin_2019}. Furthermore, the preparation of single magnon states through coupling with superconductor qubits has been demonstrated in the GHz range \cite{Lachance-Quirion20}. In principle, it is possible to realize a similar concept in the THz range by employing high Tc superconductors \cite{Sanchez-Manzano2022}. Thus it could become feasible to transfer quantum non-linearities into hybrid-magnon-quantum well polariton systems in the ultra-strong coupling regime at 1 THz, overcoming the intrinsic limitations owed to the bosonic character of these systems. We can also benefit from technologies available at this frequency range, such as semiconductor optoelectronics and THz-time domain to conceive original quantum optical experiments.  This is another venue for bringing quantum technologies into the THz frequency range \cite{TodorovDhillonMangeney2024, RevUSC_2_2019}, and paving the way towards innovative tunable THz devices operating at room temperature.

\begin{acknowledgments}
We acknowledge the funding from Pasific2 of the Polish Academy of Sciences sponsored by the European Union’s Horizon 2020 research and innovation program under the Marie Skłodowska-Curie grant agreement No.\ 847639 and from the Ministry of Education and Science of Poland. This work was also partially supported by the “International Research Agendas” program of the Foundation for Polish Science, co-financed by the European Union under the European Regional Development Fund (No.\ MAB/2018/9). We also acknowledge funding from  ERC-COG-863487 “UNIQUE”.
\end{acknowledgments}

\appendix

\section{Quantum model}\label{AnnexQM}
\begin{figure}
    \centering
    \includegraphics[width=.8\linewidth]{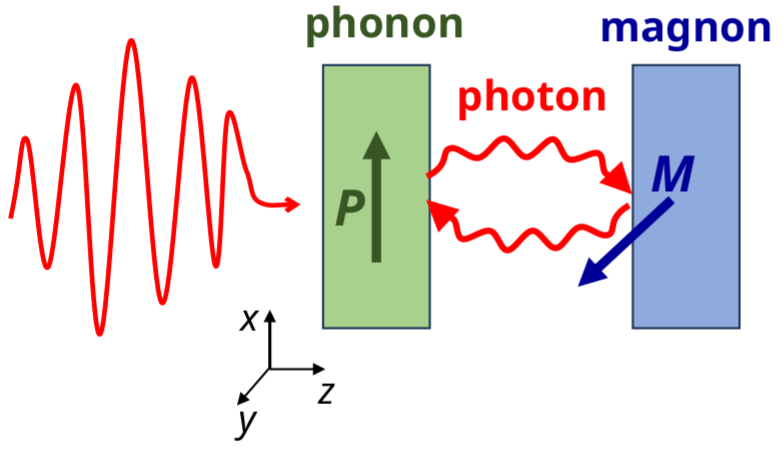}
    \caption{Schematic of the tripartite system composed of phonon, magnon, and mediating cavity photon.}
    \label{Tripartite}
\end{figure}
This Appendix provides an effective quantum model for the interacting tripartite system (Fig.\  \ref{Tripartite}). We have two objectives: the first one is to provide a quantum Hamiltonian that is compatible with effective dielectric and magnetic functions, where the polariton frequencies derived from the model can be directly compared with the data. Our second objective is to provide analytical expressions for the Hopfield mixing coefficients that allow evaluating the mixing between magnons and phonons mediated by the electromagnetic field.

\subsection{Full Hamiltonian and photon field}

We start with a very general Hamiltonian which describes photons coupled both to electrically and magnetically active matter:

\begin{eqnarray}
\mathcal{\hat{H}} = \mathcal{\hat{H}}_{phonon}+ \mathcal{\hat{H}}_{magnon} \nonumber
\\ +\int_V \Bigg[ \frac{\hat{\mathbf{D}}^2(\mathbf{r})}{2\varepsilon_{\infty}(z) \varepsilon_0} + \frac{\mu_0\hat{\mathbf{H}}^2(\mathbf{r})}{2 }\Bigg] d^3\mathbf{r}   \nonumber
\\
-\int_V \frac{\hat{\mathbf{D}}(\mathbf{r})\hat{\mathbf{P}}(\mathbf{r})}{\varepsilon_{\infty}(z) \varepsilon_0} d^3\mathbf{r} + \int_V  \frac{\hat{\mathbf{P}}^2(\mathbf{r})}{2\varepsilon(z) \varepsilon_0}d^3\mathbf{r} \nonumber
\\
-\int_V \mu_0 \hat{\mathbf{H}}(\mathbf{r})\hat{\mathbf{M}}(\mathbf{r}) d^3\mathbf{r} \label{H1}
\end{eqnarray}

The light-matter interaction Hamiltonian is expressed here in the Power-Zienau-Wooley representation of Quantum Electrodynamics \cite{Book_Cohen_Ph_At}. The first line contains the Hamiltonians of the phonon and magnon excitations that will be detailed further.  The second line of Eq. (\ref{H1}) describes the energy of the electromagnetic field. The third line describes the light-matter interaction with the dipolar excitations of interest, which in the present case are the phonons of the CBO slab. Respectively, the field $\hat{\mathbf{P}}(\mathbf{r})$ is the polarization field of the CBO phonons. In that case the dielectric function $\varepsilon_{\infty}(z)$ is a piecewise constant function that does not contain the phonon contribution and is expressed as:

\begin{eqnarray}
   \varepsilon_{\infty}(z) = \varepsilon^{CBO}_\infty, \phantom{Q} z\in CBO, \nonumber
   \\
   = \varepsilon^{NiO}_\infty,\phantom{Q} z\in NiO \nonumber
   \\
   =1, \phantom{Q} z\in \mathrm{Elsewhere}  \label{ep_func}
\end{eqnarray}

The last line of Eq. (\ref{H1}) describes the light-matter interaction with the $\alpha$ magnon of NiO \cite{Rezende19}, with $\hat{\mathbf{M}}(\mathbf{r})$ the corresponding magnetization that is provided explicitly further. 

We will consider the case where the matter systems interact with an electromagnetic wave that is polarized along the dielectric layers, along the unit vector $\mathbf{u}_x$, and propagates in the $z$-direction. We can then write a general expression of the electromagnetic field:

\begin{eqnarray}
   \hat{\mathbf{H}}(\mathbf{r})=\mathbf{u}_y \sqrt{\frac{\hbar \omega_m \varepsilon_0 c^2}{2SI}} h(z) (a+a^\dagger) \label{deffH}
   \\
  \hat{\mathbf{D}}(\mathbf{r})= -\mathbf{u}_x \sqrt{\frac{\hbar  \varepsilon_0 c^2}{2\omega_m SI}} \partial_z h(z) i(a-a^\dagger) \label{deffD}
  \\
  I = \int h(z)^2 dz \label{deffI}
\end{eqnarray}

Here $\omega_m$ is the frequency of the photon, $c$ is the speed of light in vacuum, $S$ is the surface of the system in the $xy$ plane, and $h(z)$ is a function that describes the solution of Maxwell's equation for a wave that propagates in the multilayered medium described by the piecewise function from Eq. (\ref{ep_func}). The quantity $I$ defined in Eq. (\ref{deffI}) is a normalization coefficient that allows the definition of the rising and lowering bosonic operators $a$ and $a^\dagger$.  We assume that we can define a finite volume of space containing the system and that contains the majority of the electromagnetic energy. The corresponding length along the $z$ direction will be noted $L$.
This assumption is questionable for the type of waveguide modes that are used in the experiment, however, we will see further how this formalism can still be applied by introducing effective overlap factors and linking the polariton dispersion to the effective dielectric and magnetic susceptibilities of the layers. Finally, in accordance with Maxwell's equation the function $h(z)$ must obey the identity:

\begin{equation}
    \int_L \frac{(\partial_z h(z))^2}{\varepsilon_\infty(z)} dz = \frac{\omega_m^2}{c^2} \int_L h(z)^2 dz \label{hid}
\end{equation}

This identity states that the total magnetic energy is equal to the total electric energy in the system, as expected for guided modes. 

\subsection{Light-phonon coupling in the CBO slab}

A very general expression for the phonon Hamiltonian and polarization is discussed in the Ref\cite{Rousseaux_PhysRevB125417_2023}. In particular, it can be shown for the optical phonon the following results hold, in the long wavelength limit:

\begin{eqnarray}
\mathcal{\hat{H}}_{phonon} + \int_{V_2}  \frac{\hat{\mathbf{P}}^2(\mathbf{r})}{2\varepsilon_\infty(z) \varepsilon_0}d^3\mathbf{r} 
= \sum_{\mathbf{Q}, \mathbf{n}} \hbar \omega_{LO} d^\dagger_{\mathbf{Q}\mathbf{n}}d_{\mathbf{Q}\mathbf{n}}
\\
\hat{\mathbf{P}}(\mathbf{r}) =
\nonumber \\ \sum_{\mathbf{Q}, \mathbf{n}} \sqrt{\frac{\hbar \varepsilon^{CBO}_\infty \varepsilon_0 (\omega_{LO}^2-\omega_{TO}^2 )}{2\omega_{LO} V_2}} \mathbf{n} e^{i\mathbf{Q}\mathbf{r}} (d_{\mathbf{Q}\mathbf{n}}+d^\dagger_{-\mathbf{Q}\mathbf{n}}) \label{phononPol1}
\end{eqnarray}

The phonon excitations are expanded in a plane-wave basis of bosonic basis set described by the operators $d_{\mathbf{Q}\mathbf{n}}$ and $d^\dagger_{\mathbf{Q}\mathbf{n}}$
, where the $\mathbf{Q}$ is the propagation wavevector and $\mathbf{n}$ is a unit vector defined as $\mathbf{n.Q}=0$ which fixes the polarization of the phonon wave. Here $V_2$ is the volume of the CBO slab of thickness $\Delta_2$, and the two phonon frequencies satisfy the Lyddane–Sachs–Teller relation:

\begin{equation}
\varepsilon^{CBO}_s \omega_{TO}^2 =\varepsilon^{CBO}_\infty \omega_{LO}^2
\end{equation}

The plane waves are normalized such as:

\begin{equation}
\int_{V_2} e^{i\mathbf{Q}\mathbf{r}} d^3 \mathbf{r} = V_2 \delta_{\mathbf{Q}\mathbf{0}} \label{normcondV2}
\end{equation}

We use Eq.(\ref{phononPol1}) together with  Eq.(\ref{deffD}) to express the light-matter coupling term, where the integral runs over the volume of the CBO slab alone:

\begin{eqnarray}
-\int_{V_2} \frac{\hat{\mathbf{D}}(\mathbf{r})\hat{\mathbf{P}}(\mathbf{r})}{\varepsilon^{CBO}_\infty \varepsilon_0} d^3\mathbf{r} =\nonumber \\
i(a-a^\dagger)\frac{\hbar c}{2 \sqrt{\varepsilon^{CBO}_\infty}} \sqrt{\frac{\omega_{LO}^2-\omega_{TO}^2 }{\omega_{LO} \omega_m}}\sqrt{\frac{1}{I\Delta_2}}\times \nonumber \\
\sum_{Q_z} \int_{\Delta_2} \partial_z h(z) e^{iQ_zz} dz (d_{Q_z\mathbf{u}_x}+ d^\dagger_{-Q_z\mathbf{u}_x})
\end{eqnarray}

Here we used the normalization condition Eq. (\ref{normcondV2}). We then introduce new bosonic operators for the phonons which integrate over all wavevectors:

\begin{eqnarray}
d = \sum_{Q_z} \int_{\Delta_2} \partial_z h(z) e^{iQ_zz} dz d_{Q_z\mathbf{u}_x} \mathcal{N}
\\
d^\dagger = \sum_{Q_z} \int_{\Delta_2} \partial_z h(z) e^{-iQ_zz} dz d_{Q_z\mathbf{u}_x} \mathcal{N}
\\
\mathcal{N}^{-1} = \sqrt{\int_{V2} (\partial h(z))^2 dz }
\end{eqnarray}

The normalization factor in the last equation ensures that the new bosonic operators satisfy $[d,d^\dagger] =1$. Next, we use Eq.(\ref{hid}) to express the integral over the function $(\partial h(z))^2$ that appears above. For that, we make an approximation that the balance between electric and magnetic energy is also satisfied within the CBO slab alone:

\begin{equation}
    \int_{\Delta_2} \frac{(\partial_z h(z))^2}{\varepsilon^{CBO}_\infty} dz \approx \frac{\omega_m^2}{c^2} \int_{\Delta_2} h(z)^2 dz 
\end{equation}

With that assumption, the light-phonon coupling Hamiltonian becomes:

\begin{eqnarray}
-\int_{V_2} \frac{\hat{\mathbf{D}}(\mathbf{r})\hat{\mathbf{P}}(\mathbf{r})}{\varepsilon^{CBO}_\infty \varepsilon_0} d^3\mathbf{r} =\nonumber \\
i\hbar \Omega_P (a-a^\dagger)(d+d^\dagger)
\\
\hbar \Omega_P = \eta_p \frac{1}{2}\hbar \sqrt{\omega_{LO}^2-\omega_{TO}^2}\sqrt{\frac{\omega_m}{\omega_{LO}}} \label{defOmP}
\\
\eta_p =  \sqrt{ \int_{\Delta_2} h(z)^2 dz \Big/ \int_L h(z)^2 dz} \label{eta_2}
\end{eqnarray}

In the above expression, we introduced a dimensionless overlap factor $0<\eta_p<1$ that quantifies the phonon coupling with the guided waves. Following Ref. \cite{Rousseaux_PhysRevB125417_2023} we can further show how we can derive a dielectric susceptibility for the CBO slab that includes the phonon contribution. We will come back to this topic in the last section of this Appendix.  

\subsection{Light-magnon coupling in the NiO slab}

In order to express the magnetization field in the NiO slab we use the theoretical description of magnons described in the review paper Ref. \cite{Rezende19}. In this section, we will use the notations from that reference. Magnons of NiO are  treated in the framework of easy-plane antiferromagnets in the AF phase (section D in Ref. \cite{Rezende19}) that are described by the following Hamiltonian:

\begin{equation}
 \mathcal{\hat{H}}_{magnon} = \sum_k [\hbar \omega_{\alpha k} \alpha_k^\dagger \alpha_k + \hbar \omega_{\beta k} \beta_k^\dagger \beta_k ] \label{Hmagnon}
\end{equation}

where $k$ is a tri-dimensional excitation vector of the system, and $\gamma = g\mu_B/\hbar$ is the gyromagnetic ratio, and $\mu_B$ is the Bohr magneton. In the long-wavelength limit, the frequencies of the two magnon modes appearing in the above Hamiltonian are provided by:

\begin{eqnarray}
 \omega_{\alpha,0}   \approx \gamma \sqrt{2H_E H_{Ax}}
 \\
 \omega_{\beta,0}  \approx \gamma \sqrt{2H_E H_{Az}}
\end{eqnarray}

Here we introduced the anisotropy fields of NiO which satisfy:

\begin{equation}
    H_E \gg H_{Ax} \gg H_{Az} \label{Hfields}
\end{equation}

Consequently, the highest frequency mode is the $\alpha$ mode which is the one observed around 1 THz in the experiments described in the main text. 

Using the inequalities (\ref{Hfields}) as well as the long-wavelength approximation we can use the results from Ref. \cite{Rezende19} in order to express the total spin along the $y$-direction of the $\alpha$-mode, as well as the corresponding magnetization vector as a function of the bosonic operators $\alpha_k^\dagger$ and $\alpha_k$:

\begin{equation}
M_y(\mathbf{r})  = -\frac{g\mu_B}{2}\Big(\frac{H_{Ax}}{H_E}\Big)^{1/4}\frac{N}{V_1} \sum_{k} \frac{e^{i\mathbf{kr}}}{\sqrt{N}}i(\alpha_k - \alpha^\dagger_{-k}) \label{magnetization}
\end{equation}

The $\beta$ mode does not contribute to this expression. Here we introduced the volume $V_1 = S\Delta_1$ of the NiO slab. The wavevectors are normalized here with respect to the number of spins $N$ in the spin sublattice:

\begin{equation}
    \sum_i e^{i\mathbf{kr}_i} = N\delta_{\mathbf{k0}}
\end{equation}

Using Eq.(\ref{magnetization}) together with Eq. (\ref{deffH}) in order to express the last line in Eq. (\ref{H1}) we have:

\begin{eqnarray}
-\int_V \mu_0 \hat{\mathbf{H}}(\mathbf{r})\hat{\mathbf{M}}(\mathbf{r}) d^3\mathbf{r}  = \nonumber \\
i\hbar \Omega_M (c-c^\dagger)(a+a^\dagger)
\\
\hbar\Omega_M = \eta_m \frac{g \mu_B}{2}\sqrt{\frac{\hbar \omega_m \mu_0 N}{2V_1}} \label{OmM}
\\
\eta_m = \sqrt{ \int_{\Delta_1} h(z)^2 dz \Big/ \int_L h(z)^2 dz} \label{eta_1}
\end{eqnarray}

Similar results have been obtained by Boventer et al. \cite{Boventer_PhysRevApplied014071_2023}. In the above expression, we introduced the collective bosonic operators:

\begin{eqnarray}
    c= \frac{1}{\sqrt{\Delta_1 \int_{\Delta_1} h(z)^2 dz}} \sum_k h(z)e^{-ik_zz}dz \alpha_k
\end{eqnarray}

that were derived in a very analogous manner to the collective phonon operators from the previous section. As a consequence, we obtain the expression of the overlap coefficient between photons and magnons $\eta_m$ (Eq. (\ref{eta_1})) that is defined in a similar manner as the overlap coefficient $\eta_p$ (Eq. (\ref{eta_2})) between photons and phonon in the CBO slab.

\subsection{Hopfield diagonalization} 

The full light-matter coupling Hamiltonian now becomes:

\begin{eqnarray}
\mathcal{H} = \hbar \omega_m a^\dagger a + \hbar \omega_{LO} d^\dagger d + \hbar \omega_{\alpha 0} c^\dagger c \nonumber \\
+i\hbar \Omega_P (a- a^\dagger)(d+d^\dagger) \nonumber \\
+i\hbar \Omega_M (a+ a^\dagger)(c-c^\dagger)
\end{eqnarray}

This is a bosonic Hamiltonian for three coupled harmonic oscillators, where photons and magnons are both coupled to the photon of a frequency $\omega_m$. In order to find the coupled modes of the system we will use the Hopfield-Bogoliubov approach [Ref], where we introduce a mixed polariton bosonic operator:

\begin{equation}
 \Pi = x a + y a^\dagger + m d+ td^\dagger + nc+sc^\dagger   
\end{equation}

The coupled mode frequencies $\omega$ and the boson operators $\Pi$ must satisfy the commutator relation $[\Pi, \mathcal{H}] =  \hbar \omega \Pi$. The latter can be cast in a matrix form:

\begin{equation}
    \mathbf{M}\mathbf{V} = \omega \mathbf{V}
\end{equation}

Here $\mathbf{V}^T = (x,y,m,t,n,s)$ is the vector of Hopfield coefficients, and the matrix $\mathbf{M}$ is provided by:

\begin{equation}
\mathbf{M} =
\begin{bmatrix}
\omega_m & 0 & i\Omega_P & -i\Omega_P &  -i\Omega_M & -i\Omega_M \\
0 & -\omega_m & -i\Omega_P & i\Omega_P &  -i\Omega_M & -i\Omega_M\\
 -i\Omega_P &  -i\Omega_P & \omega_{LO} & 0  & 0 & 0\\
 -i\Omega_P &  -i\Omega_P & 0 & -\omega_{LO} & 0 & 0 \\
 i\Omega_M &  -i\Omega_M & 0 & 0  & \omega_{\alpha0} & 0 \\
 -i\Omega_M &  i\Omega_M & 0 & 0  &  0 & -\omega_{\alpha0} 
\end{bmatrix}   
\end{equation}

Furthermore, the eigenvectors  must satisfy the normalization condition:

\begin{equation}
|x|^2 - |y|^2 + |m|^2 - |t|^2 + |n|^2 - |s|^2 = 1
\end{equation}

which stems from the requirement $[\Pi, \Pi^\dagger]= 1$. We thus can define the coefficients that describe respectively the photon, phonon, and magnon weight in the polariton states:

\begin{eqnarray}
    W_{photon} = |x|^2 - |y|^2 \\ 
    W_{phonon} = |m|^2 - |t|^2 \\
    W_{magnon} = |n|^2 - |s|^2
\end{eqnarray}

The matrix problem stated above can be solved numerically, thus providing eigenfrequencies and the Hopfield coefficients. These results have been illustrated in Fig. 5 of the main text.
Actually, thanks to the particular form of the matrix the equation providing the polariton frequencies can be provided explicitly. This equation, which states $\mathrm{det}|\mathbf{M}-\omega \mathbf{I}|=0$ is obtained to be:

\begin{gather}
    \Big( \omega_m^2 - \omega^2 - \frac{4\Omega_M^2\omega_{\alpha0}\omega_m}{\omega_{\alpha0}^2-\omega^2} \Big)\Big( \omega_m^2 - \omega^2 - \frac{4\Omega_P^2\omega_{LO}\omega_m}{\omega_{LO}^2-\omega^2} \Big) \nonumber \\
    = \frac{16\Omega_M^2\Omega_P^2 \omega_{\alpha0}\omega_{LO} \omega^2}{(\omega_{\alpha0}^2-\omega^2)(\omega_{LO}^2-\omega^2)}
    \label{eigenmodes}
\end{gather}

This equation is used to fit the experimental data. It is interesting to consider a case where $\omega_m \gg \omega_{\alpha0}, \omega_{LO}$. In that case the first line of the above equation simplifies to $\omega^4$ and we can write an effective coupled equation:

\begin{equation}
 (\omega_{\alpha0}^2-\omega^2)(\omega_{LO}^2-\omega^2) \approx    \frac{16\Omega_M^2\Omega_P^2 \omega_{\alpha0}\omega_{LO} \omega^2}{\omega_m^4} 
\end{equation}

This equation illustrates that the photon mode introduces and effective coupling between the phonon and magnon excitation. 

\subsection{Effective dielectric and magnetic susceptibilities}

Let us now discuss how the coupled mode equation Eq.(\ref{eigenmodes}) leads to effective susceptibilities. First, we consider the case where waves propagate only in the CBO slab (we can extend the thickness of this slab to infinity), such as $\eta_p=1$ and $\Omega_m = 0$. The coupled wave equation thus becomes, using Eqs. (\ref{defOmP}) and (\ref{eigenmodes}):

\begin{equation}
    \omega_m^2 - \omega^2 - \frac{(\omega_{LO}^2-\omega_{TO}^2)\omega_m^2}{\omega_{LO}^2-\omega^2} =0
\end{equation}

This equation can be recast in the form:

\begin{equation}
  \varepsilon^{CBO}_\infty \omega_m^2 = \varepsilon (\omega) \omega^2 \label{dispCBO}
\end{equation}

where we introduced the dielectric function $\varepsilon (\omega)$ of the homogeneous CBO slab provided by:

\begin{equation}
    \frac{\varepsilon^{CBO}_\infty}{\varepsilon (\omega)} =  1 + \frac{\omega_{LO}^2-\omega_{TO}^2}{\omega^{LO}-\omega^2} \label{defepCBO}
\end{equation}

This is exactly the dielectric susceptibility for a medium with an optical phonon. Indeed, by posing $\omega_m = c|\mathbf{Q}|/\sqrt{\varepsilon^{CBO}_\infty}$ Eq.(\ref{dispCBO}) becomes the Helmholtz dispersion relation for waves propagating in a CBO medium with a wavector $\mathbf{Q}$ \cite{Todorov_PhysRevB075115_2014}.

Likewise, we consider an infinite NiO slab, where the Helmholtz relation becomes:

\begin{eqnarray}
 \omega_m^2 = \mu(\omega)\omega^2   
\end{eqnarray}

By setting $\Omega_P=0$ and $\eta_m = 1$ in Eq.(\ref{OmM}) from Eq. (\ref{eigenmodes}) we obtain:

\begin{eqnarray}
    \frac{1}{\mu (\omega)} =  1 + \frac{W_A^2}{\omega_{\alpha0}^2-\omega^2} \label{defmuNiO}
    \\
W_A^2 = \omega_{\alpha0}^2 \frac{n \mu_0 \mu_B g}{2\sqrt{2} H_E}   
\end{eqnarray}

Here we introduced the density of spins $n=N/V_2$. Clearly, the above expressions establish the magnetic susceptibility of the NiO material as a function of microscopic quantities. 

We know the material susceptibilities from Eqs. (\ref{defepCBO}) and (\ref{defmuNiO}) together with the full polariton equation Eq.(\ref{eigenmodes}) in order to define effective susceptibilites. The latter is defined such as Eq.(\ref{eigenmodes}) can be rewritten in the form:

\begin{equation}
    \varepsilon^{CBO}_\infty \omega_m^2= \varepsilon_{eff} (\omega)\mu_{eff}(\omega) \omega^2
\end{equation}

After some algebra the following functions are obtained:

\begin{eqnarray}
    \frac{1}{\mu_{eff}(\omega)} = 1 + \eta_m^2 \frac{W_A^2}{\omega_{\alpha0}^2-\omega^2}
    \\
    \frac{1}{\varepsilon_{eff}(\omega)} = \frac{1}{\varepsilon^{CBO}_\infty } \Big[ 1 + \eta_p^2 \frac{(\omega_{LO}^2-\omega_{TO}^2)\omega_m^2}{\omega_{LO}^2-\omega^2} \Big]
\end{eqnarray}

The definition of $\mu_{eff}(\omega)$ is acceptable, as the only magnetic contribution arises from the NiO layer. However, the definition of $\varepsilon_{eff}(\omega)$ is problematic, as even after the removal of the CBO layer by setting $\eta_p=0$ we still have a medium with a dielectric constant $\varepsilon^{CBO}_\infty$, which ignores the presence of NiO and air dielectric layers. The following definition solves this problem:

\begin{eqnarray}
  \frac{1}{\varepsilon_{eff}(\omega)} = \frac{1}{\bar{\varepsilon}_\infty } \Big[ 1 + \bar{\eta}_p^2 \frac{(\omega_{LO}^2-\omega_{TO}^2)\omega_m^2}{\omega_{LO}^2-\omega^2} \Big]  
\\
\frac{1}{\bar{\varepsilon}_\infty } = 1- \eta_p^2 - \eta_m^2 + \frac{\eta_m^2}{\varepsilon_{NiO}^\infty} + \frac{\eta_p^2}{\varepsilon^{CBO}_\infty}
\\
\bar{\eta}_p^2 = \eta_p^2 \frac{\bar{\varepsilon}_\infty}{\varepsilon^{CBO}_\infty}
\end{eqnarray}

It is thus sufficient to use the re-normalized value $\bar{\eta}_p$ for the overlap factor in the light-phonon coupling constant defined in Eq.(\ref{defOmP}) instead of $\eta_p$ in order to obtain a polariton dispersion relation that takes into account all layers in the system. 

\section{Dielectric characterization of samples}
\label{dielectric_characterisation}
\begin{figure}
\includegraphics[width=\linewidth]{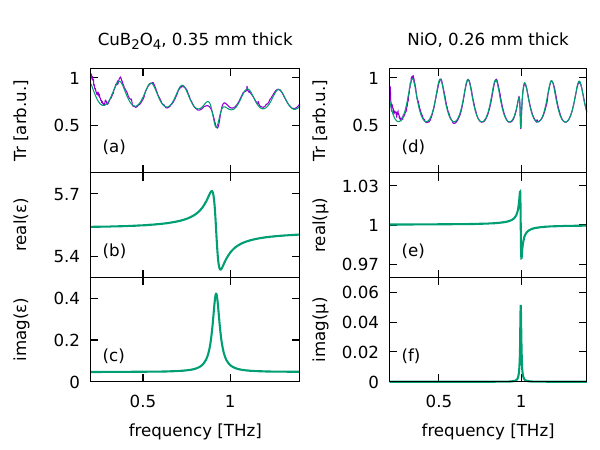}
\caption{\label{Tr}
 Transmission spectra of a standalone CBO with fits (a), under the assumption of a dielectric function with real (b) and imaginary (c) parts, permeability was assumed to be $1$. 
 Transmission spectra of stand-alone NiO with fitted transmission (d), under the assumption of permeability with real (e) and imaginary (f) parts, permittivity was assumed constant. 
}
\end{figure}

We measured transmission through each of our samples placed on a metal aperture, normalized to transmission through an empty aperture (Fig.\ \ref{Tr}). We fitted the results with a single-layer model and obtained background dielectric constants of these materials and a resonant form of permeability in the NiO crystal and phonon response in the CBO ceramics. In the case of the CBO phonon, we assumed Lorentzian lineshape; however, this model (Fig.\ \ref{Tr}bc) does not reproduce perfectly measured transmission around the resonance (Fig.\ \ref{Tr}a). This may be caused by other than the Lorentzian lineshape of the phonon mode in CBO. The parameters obtained in this way were used to calculate the predictions of electromagnetic wave reflection from structures composed of the CBO, NiO, and tunable air gap.

Thus, for our calculations, we can assume the isotropic dielectric constants of NiO and CBO, and $\epsilon_{air}=1=\mu_{air}$ for dry air. The details of the calculation methods are presented in Ref.\ \onlinecite{Bialek23}.
Here, we take into account the antiferromagnetic resonance in the permeability in NiO by writing
\begin{equation}
    \mu = 1 + \frac{\Delta\mu f_m^2}{f_m^2-f^2-ifg_m},
\end{equation}
where $f_m$ is magnon frequency, $g_m=8$ GHz is its width and its strength $\Delta\mu=4.2\times 10^{-4}$ as determined in our transmission results (Fig.\ \ref{Tr}). The permittivity of NiO was $\epsilon_{Nio}=11.63+i0.02$. The resonance strength $\Delta \mu$ is responsible for the strength of the coupling of magnon with FP modes. We take the CBO layer as a dielectric resonator
\begin{equation}
    \epsilon = \epsilon_{\infty} + \frac{\Delta\epsilon f_p^2}{f_p^2-f^2-ifg_p},
\end{equation}
where $f_p=0.92$ THz is the phonon frequency, $g_p=50$ GHz is its width, its strength $\Delta\epsilon=0.021$, and $\epsilon_{\infty}=5.52+i0.05$, as determined in our transmission results (Fig.\ \ref{Tr}). We assume $\mu_{CBO}=1$, neglecting that CBO is a paramagnet at room temperature \cite{Boehm03}.
In this electrodynamic model, we take into account the reflection from the metal mirror as a 4th layer of 100 nm thickness characterized by very high real and imaginary parts of its dielectric constant ($8\times10^8+i10^3$).

\section{Phase}\label{Phase-sign}
\begin{figure}
    \centering
    \includegraphics[width=\linewidth]{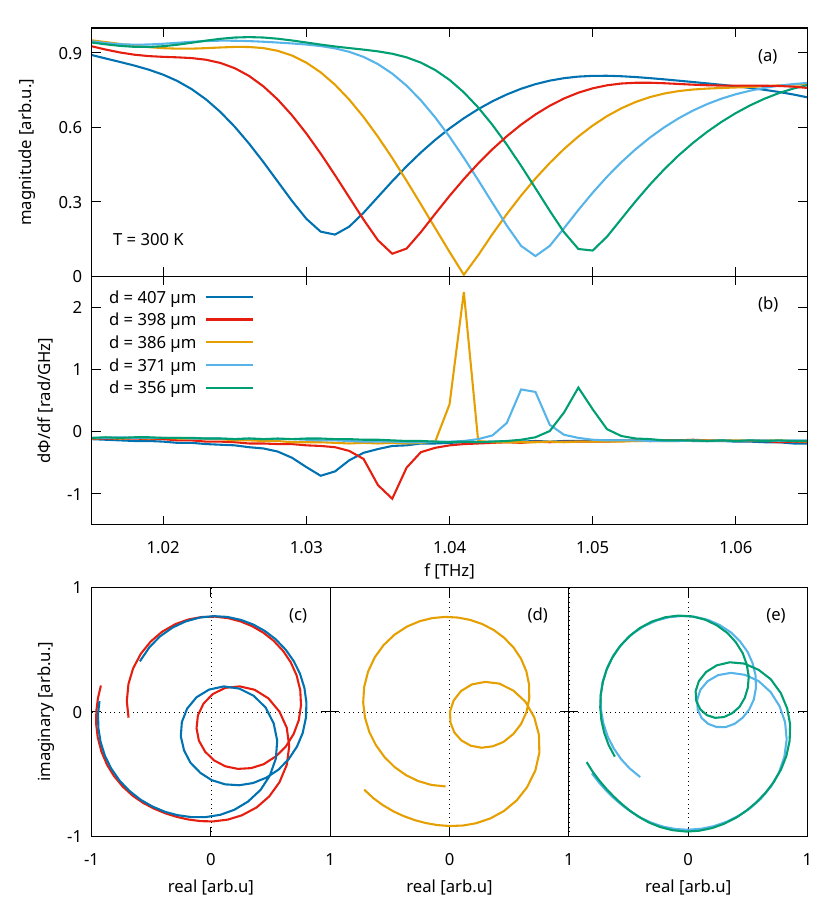}
    \caption{Evolution of a cavity mode near transition point from positive to negative phase jump. (a) amplitude and (b) phase spectra for different gaps between the slabs. Phasor plots of typical negative phase jumps encircling the origin (c), near transition condition (d), and with positive phase jumps not encircling the origin (e).}
    \label{phase-pos-neg}
\end{figure}
Here, we show in detail the evolution of a cavity mode near the transition point between positive and negative phase jump at around $f=1.04$ THz and around gap $d=400$ $\mu$m. We show in Fig.\ \ref{phase-pos-neg}a standard reflection amplitude. At larger distances of 407 and 398 $\mu$m, the cavity has a typical negative phase jumps sign (Fig.\ \ref{phase-pos-neg}b, in the direction of normal phase change with rising frequency. In Fig.\ \ref{phase-pos-neg}c, we present phasor plots for these two distances, which show that in this case, the signal circles close to the origin, reflecting the signal's lower amplitude at the cavity mode. For the case of $d=386$ $\mu$m, the transition is visible when cavity amplitude reaches almost zero and phase jump becomes undefined. Phasor trajectory for this case (Fig.\ \ref{phase-pos-neg}e) passes very close to the origin. When we further decreased the gap between the crystals to $d=371$ and 356 $\mu$m, the mode amplitude dropped again, but its phase jumps flipped sign to positive. This can be understood in phasor plot Fig.\ \ref{phase-pos-neg}e as the signal no longer circling around the origin, in a much smaller loop.

Since the data in Fig.\ \ref{phase-pos-neg} is an extract of the data presented in Fig.\ \ref{mag-pha-comparison}, we notice that the condition for the change of the phase jump sign is related to the frequency of the cavity mode, since they appear periodically for each cavity modes, when they are at the same frequency range of about 1.05 THz. Therefore this effect must be related to interference conditions, and as shown in Fig.\ \ref{phase-pos-neg} it is not related to topological changes in the electric field phasor.

\begin{figure}
    \centering
    \includegraphics[width=\linewidth]{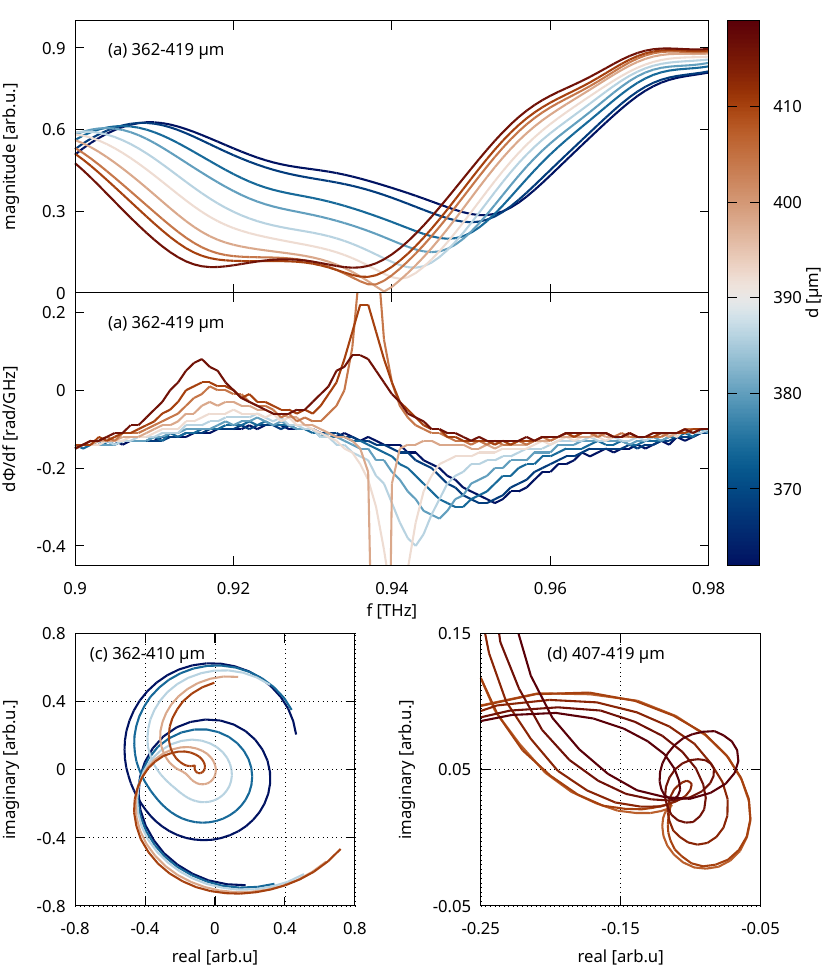}
    \caption{Evolution of phonon-polariton phasor from weak to strong coupling. (a) amplitude and (b) phase spectra for different gaps between the slabs. In the weak coupling regime, the phasor shows a single loop (c), while in the strong coupling regime, a topological change of the phasor into a double loop is observed (d).}
    \label{phase-topological}
\end{figure}
The change of mode phase jump also appears near the polariton formation regions. However, in this case not only a phase sign change is observed, but also a topological change in the phasor. In Fig.\ \ref{phase-topological}, we show in detail the evolution of phonon-polariton formation at $T=300$~K, when the magnon is away from the interaction region. Magnitude and phase spectra were collected for different gaps between the slabs, which tuned the cavity mode frequency. The strongest coupling was observed at $d=419$~$\mu$m when the cavity mode reached the phonon frequency. In Fig.\ \ref{phase-topological}c, we show phasors of a weakly interacting cavity mode, which forms a loop around the origin. The magnon can is represented as a weak deformation of the loop for higher $d$. In Fig.\ \ref{phase-topological}d, we show that a single loop is transformed into a double loop when entering the strong coupling regime, which is a topological change of the phasor.

\section{Magnon and phonon contents using electromagnetism}
Classical electrodynamics predicts some distant coupling between the resonators, which can be observed in the reflection or transmission of electromagnetic waves. Using our classical model, we determined the relative phonon and magnon contents of polariton modes. We did that by calculating the reflection amplitude derivatives to the small magnon strength change $\Delta\mu$ (1. 0\%) and the phonon strength change $\Delta\epsilon$ (0.1\%). Changes that give similar amplitude of reflection changes.
\begin{figure}
\includegraphics[width=\linewidth]{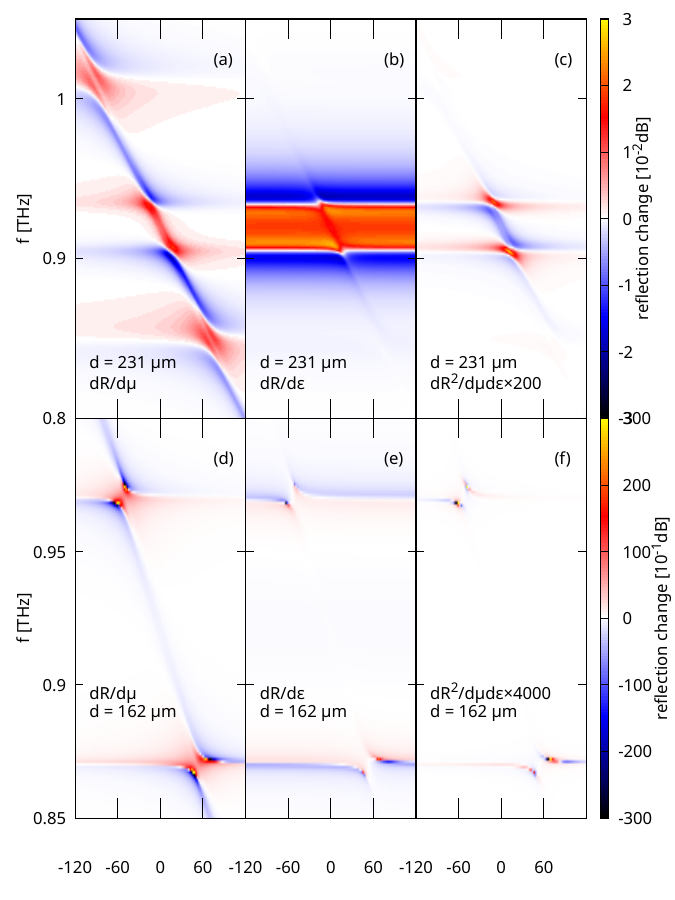}
\caption{\label{mag-phon-derivs}
 Expected magnon and phonon content, for two different gaps, one with cavity mode matching the phonon mode (a-c) and without matching (d-f). Derivative to magnon strength (a,d) shows the magnon content of cavity modes, and derivative to phonon strength (b,e) shows phonon content. The second derivative shows modes coupled simultaneously to the magnon and phonon.
}
\end{figure}
One sees that the magnon is visible in derivative to its strength (Fig.\ \ref{mag-phon-derivs}a) around the diagonal, which is magnon frequency in our model. Signatures of the magnon in reflection are either positive or negative, depending on the phase of its interaction with cavity modes. Phonon is uncovered in around its frequency in derivative to its strength (Fig.\ \ref{mag-phon-derivs}b). Here, the shape of the phonon derivative spectrum is caused by interaction with a tuned cavity mode, so that enhancing phonon strength increases the coupling rate, which results in polariton modes more separated, which is visible as higher reflection (red-yellow) in between the polaritons and lower amplitude outside (blue-black). In this phonon derivative one already sees a trace of magnon interactions, in the form of splittings of upper and lower phonon-polaritons. In particular, for $f_p-f_m=0$, the splitting of the polariton modes is slightly larger, as shown in Fig.\ \ref{mag-phon-cut} and \ref{mag-phon-cut-theo}, which is an indication of coupling of a cavity mode coherently with magnon and phonon.

However, such a coherent coupling is not the only possibility, the model shows (Fig.\ \ref{mag-phon-derivs}(d-f)) that even for modes not matching, there may be conditions in which they interact with both the magnon and the phonon. This comes from the fact that the phonon is such a strong oscillator that it couples even with cavity modes detuning as much as about 50 GHz, or more \cite{Sivarajah19}. In Fig.\ \ref{mag-phon-derivs}(d-f), one sees magnon-polariton modes that are simultaneously coupled to the phonon; however, due to a huge detuning, the coupling of these modes to the phonon is weaker. Such modes give larger differential signals, which is caused by their narrow line width, which in turn originates from their smaller phonon content than in the case of modes presented in Fig.\ \ref{mag-phon-derivs}(a-c). We observed such modes experimentally in Fig.\ \ref{pha-T}, below 0.9 THz. However, in the experiment, we cannot determine the phonon content of these modes, so we cannot show that they are indeed coupled to the phonon, in another way than in the theoretical model.

\section{Phonon-magnon coupling at large distances}
\begin{figure}
\includegraphics[width=\linewidth]{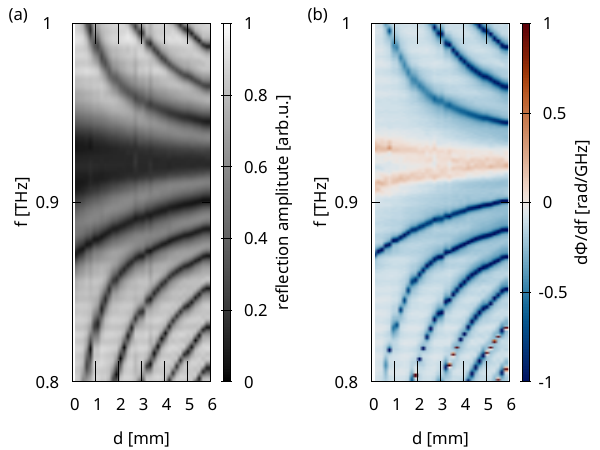}
\caption{\label{long}
 Magnitude (a) and phase differential (b) are measured as a function of the gap between the slabs with a step of 163 $\mu$m, which always gives a cavity mode as the phonon frequency. Since the results were obtained for $T=364$ K, there was also the magnon mode at the same frequency, leading to the formation of phonon-magnon polaritons. The closing of the polariton gap is visible with increasing distance $d$, up to about 4-5 mm.
}
\end{figure}
Our mechanical systems allowed to control the gap between the sample slabs in the range of tens of mm with $\mu$m precision. The focal point of the quasi-optical systems was very elongated because we used a parabolic mirror of a focal length of 101 mm. Therefore, we can widen the gap between the crystals up to a few mm without any of the samples leaving the focal point and remaining with high position accuracy. This allowed us to measure reflection for the slabs separated by large gaps $d$, which were so chosen that for each gap a cavity mode had a frequency at the phonon frequency $f_p$. The step was 163 $\mu$m, which is half a wavelength of $f_p$. This means we were stepping through subsequent cavity modes so that for each of these distances, there was a cavity mode at $f_p$, which created conditions for cavity-mediated phonon-magnon coupling. With rising distance the cavity volume $V_c$ increases, but the number of oscillators (phonons and magnons) remains constant. Since the coupling strength is proportional to $\sqrt{N/V_c}$, this explains the drop in splitting with the rising gap between the slabs. The rising gap finally leads to a collapse of the splitting, e.i.\ the transition from the strong coupling regime (polaritons) to the weak coupling regime (Purcell effect). 

\bibliography{refs}

\begin{thebibliography}{71}%
\makeatletter
\providecommand \@ifxundefined [1]{%
 \@ifx{#1\undefined}
}%
\providecommand \@ifnum [1]{%
 \ifnum #1\expandafter \@firstoftwo
 \else \expandafter \@secondoftwo
 \fi
}%
\providecommand \@ifx [1]{%
 \ifx #1\expandafter \@firstoftwo
 \else \expandafter \@secondoftwo
 \fi
}%
\providecommand \natexlab [1]{#1}%
\providecommand \enquote  [1]{``#1''}%
\providecommand \bibnamefont  [1]{#1}%
\providecommand \bibfnamefont [1]{#1}%
\providecommand \citenamefont [1]{#1}%
\providecommand \href@noop [0]{\@secondoftwo}%
\providecommand \href [0]{\begingroup \@sanitize@url \@href}%
\providecommand \@href[1]{\@@startlink{#1}\@@href}%
\providecommand \@@href[1]{\endgroup#1\@@endlink}%
\providecommand \@sanitize@url [0]{\catcode `\\12\catcode `\$12\catcode `\&12\catcode `\#12\catcode `\^12\catcode `\_12\catcode `\%12\relax}%
\providecommand \@@startlink[1]{}%
\providecommand \@@endlink[0]{}%
\providecommand \url  [0]{\begingroup\@sanitize@url \@url }%
\providecommand \@url [1]{\endgroup\@href {#1}{\urlprefix }}%
\providecommand \urlprefix  [0]{URL }%
\providecommand \Eprint [0]{\href }%
\providecommand \doibase [0]{https://doi.org/}%
\providecommand \selectlanguage [0]{\@gobble}%
\providecommand \bibinfo  [0]{\@secondoftwo}%
\providecommand \bibfield  [0]{\@secondoftwo}%
\providecommand \translation [1]{[#1]}%
\providecommand \BibitemOpen [0]{}%
\providecommand \bibitemStop [0]{}%
\providecommand \bibitemNoStop [0]{.\EOS\space}%
\providecommand \EOS [0]{\spacefactor3000\relax}%
\providecommand \BibitemShut  [1]{\csname bibitem#1\endcsname}%
\let\auto@bib@innerbib\@empty
\bibitem [{\citenamefont {Khitrova}\ \emph {et~al.}(2006)\citenamefont {Khitrova}, \citenamefont {Gibbs}, \citenamefont {Kira}, \citenamefont {Koch},\ and\ \citenamefont {Scherer}}]{Khitrova06}%
  \BibitemOpen
  \bibfield  {author} {\bibinfo {author} {\bibfnamefont {G.}~\bibnamefont {Khitrova}}, \bibinfo {author} {\bibfnamefont {H.~M.}\ \bibnamefont {Gibbs}}, \bibinfo {author} {\bibfnamefont {M.}~\bibnamefont {Kira}}, \bibinfo {author} {\bibfnamefont {S.~W.}\ \bibnamefont {Koch}},\ and\ \bibinfo {author} {\bibfnamefont {A.}~\bibnamefont {Scherer}},\ }\bibfield  {title} {\bibinfo {title} {Vacuum rabi splitting in semiconductors},\ }\href {https://doi.org/10.1038/nphys227} {\bibfield  {journal} {\bibinfo  {journal} {Nature Physics}\ }\textbf {\bibinfo {volume} {2}},\ \bibinfo {pages} {81} (\bibinfo {year} {2006})}\BibitemShut {NoStop}%
\bibitem [{\citenamefont {Dovzhenko}\ \emph {et~al.}(2018)\citenamefont {Dovzhenko}, \citenamefont {Ryabchuk}, \citenamefont {Rakovich},\ and\ \citenamefont {Nabiev}}]{Dovzhenko18}%
  \BibitemOpen
  \bibfield  {author} {\bibinfo {author} {\bibfnamefont {D.~S.}\ \bibnamefont {Dovzhenko}}, \bibinfo {author} {\bibfnamefont {S.~V.}\ \bibnamefont {Ryabchuk}}, \bibinfo {author} {\bibfnamefont {Y.~P.}\ \bibnamefont {Rakovich}},\ and\ \bibinfo {author} {\bibfnamefont {I.~R.}\ \bibnamefont {Nabiev}},\ }\bibfield  {title} {\bibinfo {title} {Light–matter interaction in the strong coupling regime: configurations{,} conditions{,} and applications},\ }\href {https://doi.org/10.1039/C7NR06917K} {\bibfield  {journal} {\bibinfo  {journal} {Nanoscale}\ }\textbf {\bibinfo {volume} {10}},\ \bibinfo {pages} {3589} (\bibinfo {year} {2018})}\BibitemShut {NoStop}%
\bibitem [{\citenamefont {Rempe}\ \emph {et~al.}(1987)\citenamefont {Rempe}, \citenamefont {Walther},\ and\ \citenamefont {Klein}}]{Rempe87}%
  \BibitemOpen
  \bibfield  {author} {\bibinfo {author} {\bibfnamefont {G.}~\bibnamefont {Rempe}}, \bibinfo {author} {\bibfnamefont {H.}~\bibnamefont {Walther}},\ and\ \bibinfo {author} {\bibfnamefont {N.}~\bibnamefont {Klein}},\ }\bibfield  {title} {\bibinfo {title} {Observation of quantum collapse and revival in a one-atom maser},\ }\href {https://doi.org/10.1103/PhysRevLett.58.353} {\bibfield  {journal} {\bibinfo  {journal} {Phys. Rev. Lett.}\ }\textbf {\bibinfo {volume} {58}},\ \bibinfo {pages} {353} (\bibinfo {year} {1987})}\BibitemShut {NoStop}%
\bibitem [{\citenamefont {Kasprzak}\ \emph {et~al.}(2006)\citenamefont {Kasprzak}, \citenamefont {Richard}, \citenamefont {Kundermann}, \citenamefont {Baas}, \citenamefont {Jeambrun}, \citenamefont {Keeling}, \citenamefont {Marchetti}, \citenamefont {Szyma{\'{n}}ska}, \citenamefont {Andr{\'e}}, \citenamefont {Staehli}, \citenamefont {Savona}, \citenamefont {Littlewood}, \citenamefont {Deveaud},\ and\ \citenamefont {Dang}}]{Kasprzak06}%
  \BibitemOpen
  \bibfield  {author} {\bibinfo {author} {\bibfnamefont {J.}~\bibnamefont {Kasprzak}}, \bibinfo {author} {\bibfnamefont {M.}~\bibnamefont {Richard}}, \bibinfo {author} {\bibfnamefont {S.}~\bibnamefont {Kundermann}}, \bibinfo {author} {\bibfnamefont {A.}~\bibnamefont {Baas}}, \bibinfo {author} {\bibfnamefont {P.}~\bibnamefont {Jeambrun}}, \bibinfo {author} {\bibfnamefont {J.~M.~J.}\ \bibnamefont {Keeling}}, \bibinfo {author} {\bibfnamefont {F.~M.}\ \bibnamefont {Marchetti}}, \bibinfo {author} {\bibfnamefont {M.~H.}\ \bibnamefont {Szyma{\'{n}}ska}}, \bibinfo {author} {\bibfnamefont {R.}~\bibnamefont {Andr{\'e}}}, \bibinfo {author} {\bibfnamefont {J.~L.}\ \bibnamefont {Staehli}}, \bibinfo {author} {\bibfnamefont {V.}~\bibnamefont {Savona}}, \bibinfo {author} {\bibfnamefont {P.~B.}\ \bibnamefont {Littlewood}}, \bibinfo {author} {\bibfnamefont {B.}~\bibnamefont {Deveaud}},\ and\ \bibinfo {author} {\bibfnamefont {L.~S.}\ \bibnamefont {Dang}},\ }\bibfield  {title} {\bibinfo {title} {Bose--einstein condensation of
  exciton polaritons},\ }\href {https://doi.org/10.1038/nature05131} {\bibfield  {journal} {\bibinfo  {journal} {Nature}\ }\textbf {\bibinfo {volume} {443}},\ \bibinfo {pages} {409} (\bibinfo {year} {2006})}\BibitemShut {NoStop}%
\bibitem [{\citenamefont {Raizen}\ \emph {et~al.}(1989)\citenamefont {Raizen}, \citenamefont {Thompson}, \citenamefont {Brecha}, \citenamefont {Kimble},\ and\ \citenamefont {Carmichael}}]{Raizen89}%
  \BibitemOpen
  \bibfield  {author} {\bibinfo {author} {\bibfnamefont {M.~G.}\ \bibnamefont {Raizen}}, \bibinfo {author} {\bibfnamefont {R.~J.}\ \bibnamefont {Thompson}}, \bibinfo {author} {\bibfnamefont {R.~J.}\ \bibnamefont {Brecha}}, \bibinfo {author} {\bibfnamefont {H.~J.}\ \bibnamefont {Kimble}},\ and\ \bibinfo {author} {\bibfnamefont {H.~J.}\ \bibnamefont {Carmichael}},\ }\bibfield  {title} {\bibinfo {title} {Normal-mode splitting and linewidth averaging for two-state atoms in an optical cavity},\ }\href {https://doi.org/10.1103/PhysRevLett.63.240} {\bibfield  {journal} {\bibinfo  {journal} {Phys. Rev. Lett.}\ }\textbf {\bibinfo {volume} {63}},\ \bibinfo {pages} {240} (\bibinfo {year} {1989})}\BibitemShut {NoStop}%
\bibitem [{\citenamefont {Colombe}\ \emph {et~al.}(2007)\citenamefont {Colombe}, \citenamefont {Steinmetz}, \citenamefont {Dubois}, \citenamefont {Linke}, \citenamefont {Hunger},\ and\ \citenamefont {Reichel}}]{Colombe07}%
  \BibitemOpen
  \bibfield  {author} {\bibinfo {author} {\bibfnamefont {Y.}~\bibnamefont {Colombe}}, \bibinfo {author} {\bibfnamefont {T.}~\bibnamefont {Steinmetz}}, \bibinfo {author} {\bibfnamefont {G.}~\bibnamefont {Dubois}}, \bibinfo {author} {\bibfnamefont {F.}~\bibnamefont {Linke}}, \bibinfo {author} {\bibfnamefont {D.}~\bibnamefont {Hunger}},\ and\ \bibinfo {author} {\bibfnamefont {J.}~\bibnamefont {Reichel}},\ }\bibfield  {title} {\bibinfo {title} {Strong atom--field coupling for bose--einstein condensates in an optical cavity on a chip},\ }\href {https://doi.org/10.1038/nature06331} {\bibfield  {journal} {\bibinfo  {journal} {Nature}\ }\textbf {\bibinfo {volume} {450}},\ \bibinfo {pages} {272} (\bibinfo {year} {2007})}\BibitemShut {NoStop}%
\bibitem [{\citenamefont {Delteil}\ \emph {et~al.}(2012)\citenamefont {Delteil}, \citenamefont {Vasanelli}, \citenamefont {Todorov}, \citenamefont {Feuillet~Palma}, \citenamefont {Renaudat St-Jean}, \citenamefont {Beaudoin}, \citenamefont {Sagnes},\ and\ \citenamefont {Sirtori}}]{Delteil_prl2012}%
  \BibitemOpen
  \bibfield  {author} {\bibinfo {author} {\bibfnamefont {A.}~\bibnamefont {Delteil}}, \bibinfo {author} {\bibfnamefont {A.}~\bibnamefont {Vasanelli}}, \bibinfo {author} {\bibfnamefont {Y.}~\bibnamefont {Todorov}}, \bibinfo {author} {\bibfnamefont {C.}~\bibnamefont {Feuillet~Palma}}, \bibinfo {author} {\bibfnamefont {M.}~\bibnamefont {Renaudat St-Jean}}, \bibinfo {author} {\bibfnamefont {G.}~\bibnamefont {Beaudoin}}, \bibinfo {author} {\bibfnamefont {I.}~\bibnamefont {Sagnes}},\ and\ \bibinfo {author} {\bibfnamefont {C.}~\bibnamefont {Sirtori}},\ }\bibfield  {title} {\bibinfo {title} {Charge-induced coherence between intersubband plasmons in a quantum structure},\ }\href {https://doi.org/10.1103/PhysRevLett.109.246808} {\bibfield  {journal} {\bibinfo  {journal} {Phys. Rev. Lett.}\ }\textbf {\bibinfo {volume} {109}},\ \bibinfo {pages} {246808} (\bibinfo {year} {2012})}\BibitemShut {NoStop}%
\bibitem [{\citenamefont {Törmä}\ and\ \citenamefont {Barnes}(2014)}]{Torma14}%
  \BibitemOpen
  \bibfield  {author} {\bibinfo {author} {\bibfnamefont {P.}~\bibnamefont {Törmä}}\ and\ \bibinfo {author} {\bibfnamefont {W.~L.}\ \bibnamefont {Barnes}},\ }\bibfield  {title} {\bibinfo {title} {Strong coupling between surface plasmon polaritons and emitters: a review},\ }\href {https://doi.org/10.1088/0034-4885/78/1/013901} {\bibfield  {journal} {\bibinfo  {journal} {Reports on Progress in Physics}\ }\textbf {\bibinfo {volume} {78}},\ \bibinfo {pages} {013901} (\bibinfo {year} {2014})}\BibitemShut {NoStop}%
\bibitem [{\citenamefont {Basov}\ \emph {et~al.}(2016)\citenamefont {Basov}, \citenamefont {Fogler},\ and\ \citenamefont {de~Abajo}}]{Basov16}%
  \BibitemOpen
  \bibfield  {author} {\bibinfo {author} {\bibfnamefont {D.~N.}\ \bibnamefont {Basov}}, \bibinfo {author} {\bibfnamefont {M.~M.}\ \bibnamefont {Fogler}},\ and\ \bibinfo {author} {\bibfnamefont {F.~J.~G.}\ \bibnamefont {de~Abajo}},\ }\bibfield  {title} {\bibinfo {title} {{Polaritons in van der Waals materials}},\ }\href {https://doi.org/10.1126/science.aag1992} {\bibfield  {journal} {\bibinfo  {journal} {Science}\ }\textbf {\bibinfo {volume} {354}},\ \bibinfo {pages} {aag1992} (\bibinfo {year} {2016})}\BibitemShut {NoStop}%
\bibitem [{\citenamefont {Bayer}\ \emph {et~al.}(2017)\citenamefont {Bayer}, \citenamefont {Pozimski}, \citenamefont {Schambeck}, \citenamefont {Schuh}, \citenamefont {Huber}, \citenamefont {Bougeard},\ and\ \citenamefont {Lange}}]{Bayer17}%
  \BibitemOpen
  \bibfield  {author} {\bibinfo {author} {\bibfnamefont {A.}~\bibnamefont {Bayer}}, \bibinfo {author} {\bibfnamefont {M.}~\bibnamefont {Pozimski}}, \bibinfo {author} {\bibfnamefont {S.}~\bibnamefont {Schambeck}}, \bibinfo {author} {\bibfnamefont {D.}~\bibnamefont {Schuh}}, \bibinfo {author} {\bibfnamefont {R.}~\bibnamefont {Huber}}, \bibinfo {author} {\bibfnamefont {D.}~\bibnamefont {Bougeard}},\ and\ \bibinfo {author} {\bibfnamefont {C.}~\bibnamefont {Lange}},\ }\bibfield  {title} {\bibinfo {title} {Terahertz light--matter interaction beyond unity coupling strength},\ }\href {https://doi.org/10.1021/acs.nanolett.7b03103} {\bibfield  {journal} {\bibinfo  {journal} {Nano Letters}\ }\textbf {\bibinfo {volume} {17}},\ \bibinfo {pages} {6340} (\bibinfo {year} {2017})}\BibitemShut {NoStop}%
\bibitem [{\citenamefont {Li}\ \emph {et~al.}(2018)\citenamefont {Li}, \citenamefont {Bamba}, \citenamefont {Yuan}, \citenamefont {Zhang}, \citenamefont {Zhao}, \citenamefont {Xiang}, \citenamefont {Xu}, \citenamefont {Jin}, \citenamefont {Ren}, \citenamefont {Ma}, \citenamefont {Cao}, \citenamefont {Turchinovich},\ and\ \citenamefont {Kono}}]{Li18}%
  \BibitemOpen
  \bibfield  {author} {\bibinfo {author} {\bibfnamefont {X.}~\bibnamefont {Li}}, \bibinfo {author} {\bibfnamefont {M.}~\bibnamefont {Bamba}}, \bibinfo {author} {\bibfnamefont {N.}~\bibnamefont {Yuan}}, \bibinfo {author} {\bibfnamefont {Q.}~\bibnamefont {Zhang}}, \bibinfo {author} {\bibfnamefont {Y.}~\bibnamefont {Zhao}}, \bibinfo {author} {\bibfnamefont {M.}~\bibnamefont {Xiang}}, \bibinfo {author} {\bibfnamefont {K.}~\bibnamefont {Xu}}, \bibinfo {author} {\bibfnamefont {Z.}~\bibnamefont {Jin}}, \bibinfo {author} {\bibfnamefont {W.}~\bibnamefont {Ren}}, \bibinfo {author} {\bibfnamefont {G.}~\bibnamefont {Ma}}, \bibinfo {author} {\bibfnamefont {S.}~\bibnamefont {Cao}}, \bibinfo {author} {\bibfnamefont {D.}~\bibnamefont {Turchinovich}},\ and\ \bibinfo {author} {\bibfnamefont {J.}~\bibnamefont {Kono}},\ }\bibfield  {title} {\bibinfo {title} {{Observation of Dicke cooperativity in magnetic interactions}},\ }\href {https://doi.org/10.1126/science.aat5162} {\bibfield  {journal} {\bibinfo  {journal} {Science}\
  }\textbf {\bibinfo {volume} {361}},\ \bibinfo {pages} {794} (\bibinfo {year} {2018})}\BibitemShut {NoStop}%
\bibitem [{\citenamefont {Yahiaoui}\ \emph {et~al.}(2022)\citenamefont {Yahiaoui}, \citenamefont {Chase}, \citenamefont {Kyaw}, \citenamefont {Tay}, \citenamefont {Baydin}, \citenamefont {Noe}, \citenamefont {Song}, \citenamefont {Kono}, \citenamefont {Agrawal}, \citenamefont {Bamba},\ and\ \citenamefont {Searles}}]{Yahiaoui22}%
  \BibitemOpen
  \bibfield  {author} {\bibinfo {author} {\bibfnamefont {R.}~\bibnamefont {Yahiaoui}}, \bibinfo {author} {\bibfnamefont {Z.~A.}\ \bibnamefont {Chase}}, \bibinfo {author} {\bibfnamefont {C.}~\bibnamefont {Kyaw}}, \bibinfo {author} {\bibfnamefont {F.}~\bibnamefont {Tay}}, \bibinfo {author} {\bibfnamefont {A.}~\bibnamefont {Baydin}}, \bibinfo {author} {\bibfnamefont {G.~T.}\ \bibnamefont {Noe}}, \bibinfo {author} {\bibfnamefont {J.}~\bibnamefont {Song}}, \bibinfo {author} {\bibfnamefont {J.}~\bibnamefont {Kono}}, \bibinfo {author} {\bibfnamefont {A.}~\bibnamefont {Agrawal}}, \bibinfo {author} {\bibfnamefont {M.}~\bibnamefont {Bamba}},\ and\ \bibinfo {author} {\bibfnamefont {T.~A.}\ \bibnamefont {Searles}},\ }\bibfield  {title} {\bibinfo {title} {Dicke-cooperativity-assisted ultrastrong coupling enhancement in terahertz metasurfaces},\ }\bibfield  {journal} {\bibinfo  {journal} {Nano Letters}\ }\href {https://doi.org/10.1021/acs.nanolett.2c01892} {10.1021/acs.nanolett.2c01892} (\bibinfo {year} {2022})\BibitemShut
  {NoStop}%
\bibitem [{\citenamefont {Roberts}\ \emph {et~al.}(1962)\citenamefont {Roberts}, \citenamefont {Auld},\ and\ \citenamefont {Schell}}]{Roberts62}%
  \BibitemOpen
  \bibfield  {author} {\bibinfo {author} {\bibfnamefont {R.~W.}\ \bibnamefont {Roberts}}, \bibinfo {author} {\bibfnamefont {B.~A.}\ \bibnamefont {Auld}},\ and\ \bibinfo {author} {\bibfnamefont {R.~R.}\ \bibnamefont {Schell}},\ }\bibfield  {title} {\bibinfo {title} {Magnetodynamic mode ferrite amplifier},\ }\href {https://doi.org/10.1063/1.1728685} {\bibfield  {journal} {\bibinfo  {journal} {Journal of Applied Physics}\ }\textbf {\bibinfo {volume} {33}},\ \bibinfo {pages} {1267} (\bibinfo {year} {1962})}\BibitemShut {NoStop}%
\bibitem [{\citenamefont {Schuster}\ \emph {et~al.}(2010)\citenamefont {Schuster}, \citenamefont {Sears}, \citenamefont {Ginossar}, \citenamefont {DiCarlo}, \citenamefont {Frunzio}, \citenamefont {Morton}, \citenamefont {Wu}, \citenamefont {Briggs}, \citenamefont {Buckley}, \citenamefont {Awschalom},\ and\ \citenamefont {Schoelkopf}}]{Schuster10}%
  \BibitemOpen
  \bibfield  {author} {\bibinfo {author} {\bibfnamefont {D.~I.}\ \bibnamefont {Schuster}}, \bibinfo {author} {\bibfnamefont {A.~P.}\ \bibnamefont {Sears}}, \bibinfo {author} {\bibfnamefont {E.}~\bibnamefont {Ginossar}}, \bibinfo {author} {\bibfnamefont {L.}~\bibnamefont {DiCarlo}}, \bibinfo {author} {\bibfnamefont {L.}~\bibnamefont {Frunzio}}, \bibinfo {author} {\bibfnamefont {J.~J.~L.}\ \bibnamefont {Morton}}, \bibinfo {author} {\bibfnamefont {H.}~\bibnamefont {Wu}}, \bibinfo {author} {\bibfnamefont {G.~A.~D.}\ \bibnamefont {Briggs}}, \bibinfo {author} {\bibfnamefont {B.~B.}\ \bibnamefont {Buckley}}, \bibinfo {author} {\bibfnamefont {D.~D.}\ \bibnamefont {Awschalom}},\ and\ \bibinfo {author} {\bibfnamefont {R.~J.}\ \bibnamefont {Schoelkopf}},\ }\bibfield  {title} {\bibinfo {title} {High-cooperativity coupling of electron-spin ensembles to superconducting cavities},\ }\href {https://doi.org/10.1103/PhysRevLett.105.140501} {\bibfield  {journal} {\bibinfo  {journal} {Phys. Rev. Lett.}\ }\textbf {\bibinfo
  {volume} {105}},\ \bibinfo {pages} {140501} (\bibinfo {year} {2010})}\BibitemShut {NoStop}%
\bibitem [{\citenamefont {Abe}\ \emph {et~al.}(2011)\citenamefont {Abe}, \citenamefont {Wu}, \citenamefont {Ardavan},\ and\ \citenamefont {Morton}}]{Abe11}%
  \BibitemOpen
  \bibfield  {author} {\bibinfo {author} {\bibfnamefont {E.}~\bibnamefont {Abe}}, \bibinfo {author} {\bibfnamefont {H.}~\bibnamefont {Wu}}, \bibinfo {author} {\bibfnamefont {A.}~\bibnamefont {Ardavan}},\ and\ \bibinfo {author} {\bibfnamefont {J.~J.~L.}\ \bibnamefont {Morton}},\ }\bibfield  {title} {\bibinfo {title} {Electron spin ensemble strongly coupled to a three-dimensional microwave cavity},\ }\href {https://doi.org/10.1063/1.3601930} {\bibfield  {journal} {\bibinfo  {journal} {Applied Physics Letters}\ }\textbf {\bibinfo {volume} {98}},\ \bibinfo {pages} {251108} (\bibinfo {year} {2011})}\BibitemShut {NoStop}%
\bibitem [{\citenamefont {Huebl}\ \emph {et~al.}(2013)\citenamefont {Huebl}, \citenamefont {Zollitsch}, \citenamefont {Lotze}, \citenamefont {Hocke}, \citenamefont {Greifenstein}, \citenamefont {Marx}, \citenamefont {Gross},\ and\ \citenamefont {Goennenwein}}]{Huebl13}%
  \BibitemOpen
  \bibfield  {author} {\bibinfo {author} {\bibfnamefont {H.}~\bibnamefont {Huebl}}, \bibinfo {author} {\bibfnamefont {C.~W.}\ \bibnamefont {Zollitsch}}, \bibinfo {author} {\bibfnamefont {J.}~\bibnamefont {Lotze}}, \bibinfo {author} {\bibfnamefont {F.}~\bibnamefont {Hocke}}, \bibinfo {author} {\bibfnamefont {M.}~\bibnamefont {Greifenstein}}, \bibinfo {author} {\bibfnamefont {A.}~\bibnamefont {Marx}}, \bibinfo {author} {\bibfnamefont {R.}~\bibnamefont {Gross}},\ and\ \bibinfo {author} {\bibfnamefont {S.~T.~B.}\ \bibnamefont {Goennenwein}},\ }\bibfield  {title} {\bibinfo {title} {High cooperativity in coupled microwave resonator ferrimagnetic insulator hybrids},\ }\href {https://doi.org/10.1103/PhysRevLett.111.127003} {\bibfield  {journal} {\bibinfo  {journal} {Phys. Rev. Lett.}\ }\textbf {\bibinfo {volume} {111}},\ \bibinfo {pages} {127003} (\bibinfo {year} {2013})}\BibitemShut {NoStop}%
\bibitem [{\citenamefont {Zhang}\ \emph {et~al.}(2014)\citenamefont {Zhang}, \citenamefont {Zou}, \citenamefont {Jiang},\ and\ \citenamefont {Tang}}]{Zhang14}%
  \BibitemOpen
  \bibfield  {author} {\bibinfo {author} {\bibfnamefont {X.}~\bibnamefont {Zhang}}, \bibinfo {author} {\bibfnamefont {C.-L.}\ \bibnamefont {Zou}}, \bibinfo {author} {\bibfnamefont {L.}~\bibnamefont {Jiang}},\ and\ \bibinfo {author} {\bibfnamefont {H.~X.}\ \bibnamefont {Tang}},\ }\bibfield  {title} {\bibinfo {title} {Strongly coupled magnons and cavity microwave photons},\ }\href {https://doi.org/10.1103/PhysRevLett.113.156401} {\bibfield  {journal} {\bibinfo  {journal} {Phys. Rev. Lett.}\ }\textbf {\bibinfo {volume} {113}},\ \bibinfo {pages} {156401} (\bibinfo {year} {2014})}\BibitemShut {NoStop}%
\bibitem [{\citenamefont {Tabuchi}\ \emph {et~al.}(2014)\citenamefont {Tabuchi}, \citenamefont {Ishino}, \citenamefont {Ishikawa}, \citenamefont {Yamazaki}, \citenamefont {Usami},\ and\ \citenamefont {Nakamura}}]{Tabuchi14}%
  \BibitemOpen
  \bibfield  {author} {\bibinfo {author} {\bibfnamefont {Y.}~\bibnamefont {Tabuchi}}, \bibinfo {author} {\bibfnamefont {S.}~\bibnamefont {Ishino}}, \bibinfo {author} {\bibfnamefont {T.}~\bibnamefont {Ishikawa}}, \bibinfo {author} {\bibfnamefont {R.}~\bibnamefont {Yamazaki}}, \bibinfo {author} {\bibfnamefont {K.}~\bibnamefont {Usami}},\ and\ \bibinfo {author} {\bibfnamefont {Y.}~\bibnamefont {Nakamura}},\ }\bibfield  {title} {\bibinfo {title} {Hybridizing ferromagnetic magnons and microwave photons in the quantum limit},\ }\href {https://doi.org/10.1103/PhysRevLett.113.083603} {\bibfield  {journal} {\bibinfo  {journal} {Phys. Rev. Lett.}\ }\textbf {\bibinfo {volume} {113}},\ \bibinfo {pages} {083603} (\bibinfo {year} {2014})}\BibitemShut {NoStop}%
\bibitem [{\citenamefont {Tabuchi}\ \emph {et~al.}(2015)\citenamefont {Tabuchi}, \citenamefont {Ishino}, \citenamefont {Noguchi}, \citenamefont {Ishikawa}, \citenamefont {Yamazaki}, \citenamefont {Usami},\ and\ \citenamefont {Nakamura}}]{Tabuchi15}%
  \BibitemOpen
  \bibfield  {author} {\bibinfo {author} {\bibfnamefont {Y.}~\bibnamefont {Tabuchi}}, \bibinfo {author} {\bibfnamefont {S.}~\bibnamefont {Ishino}}, \bibinfo {author} {\bibfnamefont {A.}~\bibnamefont {Noguchi}}, \bibinfo {author} {\bibfnamefont {T.}~\bibnamefont {Ishikawa}}, \bibinfo {author} {\bibfnamefont {R.}~\bibnamefont {Yamazaki}}, \bibinfo {author} {\bibfnamefont {K.}~\bibnamefont {Usami}},\ and\ \bibinfo {author} {\bibfnamefont {Y.}~\bibnamefont {Nakamura}},\ }\bibfield  {title} {\bibinfo {title} {Coherent coupling between a ferromagnetic magnon and a superconducting qubit},\ }\href {https://doi.org/10.1126/science.aaa3693} {\bibfield  {journal} {\bibinfo  {journal} {Science}\ }\textbf {\bibinfo {volume} {349}},\ \bibinfo {pages} {405} (\bibinfo {year} {2015})}\BibitemShut {NoStop}%
\bibitem [{\citenamefont {Zhang}\ \emph {et~al.}(2015)\citenamefont {Zhang}, \citenamefont {Zou}, \citenamefont {Zhu}, \citenamefont {Marquardt}, \citenamefont {Jiang},\ and\ \citenamefont {Tang}}]{Zhang15}%
  \BibitemOpen
  \bibfield  {author} {\bibinfo {author} {\bibfnamefont {X.}~\bibnamefont {Zhang}}, \bibinfo {author} {\bibfnamefont {C.-L.}\ \bibnamefont {Zou}}, \bibinfo {author} {\bibfnamefont {N.}~\bibnamefont {Zhu}}, \bibinfo {author} {\bibfnamefont {F.}~\bibnamefont {Marquardt}}, \bibinfo {author} {\bibfnamefont {L.}~\bibnamefont {Jiang}},\ and\ \bibinfo {author} {\bibfnamefont {H.~X.}\ \bibnamefont {Tang}},\ }\bibfield  {title} {\bibinfo {title} {Magnon dark modes and gradient memory},\ }\href {https://doi.org/10.1038/ncomms9914} {\bibfield  {journal} {\bibinfo  {journal} {Nature Communications}\ }\textbf {\bibinfo {volume} {6}},\ \bibinfo {pages} {8914} (\bibinfo {year} {2015})}\BibitemShut {NoStop}%
\bibitem [{\citenamefont {Zhang}\ \emph {et~al.}(2016)\citenamefont {Zhang}, \citenamefont {Zhu}, \citenamefont {Zou},\ and\ \citenamefont {Tang}}]{Zhang16}%
  \BibitemOpen
  \bibfield  {author} {\bibinfo {author} {\bibfnamefont {X.}~\bibnamefont {Zhang}}, \bibinfo {author} {\bibfnamefont {N.}~\bibnamefont {Zhu}}, \bibinfo {author} {\bibfnamefont {C.-L.}\ \bibnamefont {Zou}},\ and\ \bibinfo {author} {\bibfnamefont {H.~X.}\ \bibnamefont {Tang}},\ }\bibfield  {title} {\bibinfo {title} {Optomagnonic whispering gallery microresonators},\ }\href {https://doi.org/10.1103/PhysRevLett.117.123605} {\bibfield  {journal} {\bibinfo  {journal} {Phys. Rev. Lett.}\ }\textbf {\bibinfo {volume} {117}},\ \bibinfo {pages} {123605} (\bibinfo {year} {2016})}\BibitemShut {NoStop}%
\bibitem [{\citenamefont {Li}\ \emph {et~al.}(2019)\citenamefont {Li}, \citenamefont {Polakovic}, \citenamefont {Wang}, \citenamefont {Xu}, \citenamefont {Lendinez}, \citenamefont {Zhang}, \citenamefont {Ding}, \citenamefont {Khaire}, \citenamefont {Saglam}, \citenamefont {Divan}, \citenamefont {Pearson}, \citenamefont {Kwok}, \citenamefont {Xiao}, \citenamefont {Novosad}, \citenamefont {Hoffmann},\ and\ \citenamefont {Zhang}}]{Li19}%
  \BibitemOpen
  \bibfield  {author} {\bibinfo {author} {\bibfnamefont {Y.}~\bibnamefont {Li}}, \bibinfo {author} {\bibfnamefont {T.}~\bibnamefont {Polakovic}}, \bibinfo {author} {\bibfnamefont {Y.-L.}\ \bibnamefont {Wang}}, \bibinfo {author} {\bibfnamefont {J.}~\bibnamefont {Xu}}, \bibinfo {author} {\bibfnamefont {S.}~\bibnamefont {Lendinez}}, \bibinfo {author} {\bibfnamefont {Z.}~\bibnamefont {Zhang}}, \bibinfo {author} {\bibfnamefont {J.}~\bibnamefont {Ding}}, \bibinfo {author} {\bibfnamefont {T.}~\bibnamefont {Khaire}}, \bibinfo {author} {\bibfnamefont {H.}~\bibnamefont {Saglam}}, \bibinfo {author} {\bibfnamefont {R.}~\bibnamefont {Divan}}, \bibinfo {author} {\bibfnamefont {J.}~\bibnamefont {Pearson}}, \bibinfo {author} {\bibfnamefont {W.-K.}\ \bibnamefont {Kwok}}, \bibinfo {author} {\bibfnamefont {Z.}~\bibnamefont {Xiao}}, \bibinfo {author} {\bibfnamefont {V.}~\bibnamefont {Novosad}}, \bibinfo {author} {\bibfnamefont {A.}~\bibnamefont {Hoffmann}},\ and\ \bibinfo {author} {\bibfnamefont {W.}~\bibnamefont {Zhang}},\
  }\bibfield  {title} {\bibinfo {title} {Strong coupling between magnons and microwave photons in on-chip ferromagnet-superconductor thin-film devices},\ }\href {https://doi.org/10.1103/PhysRevLett.123.107701} {\bibfield  {journal} {\bibinfo  {journal} {Phys. Rev. Lett.}\ }\textbf {\bibinfo {volume} {123}},\ \bibinfo {pages} {107701} (\bibinfo {year} {2019})}\BibitemShut {NoStop}%
\bibitem [{\citenamefont {Everts}\ \emph {et~al.}(2020)\citenamefont {Everts}, \citenamefont {King}, \citenamefont {Lambert}, \citenamefont {Kocsis}, \citenamefont {Rogge},\ and\ \citenamefont {Longdell}}]{Everts20}%
  \BibitemOpen
  \bibfield  {author} {\bibinfo {author} {\bibfnamefont {J.~R.}\ \bibnamefont {Everts}}, \bibinfo {author} {\bibfnamefont {G.~G.~G.}\ \bibnamefont {King}}, \bibinfo {author} {\bibfnamefont {N.~J.}\ \bibnamefont {Lambert}}, \bibinfo {author} {\bibfnamefont {S.}~\bibnamefont {Kocsis}}, \bibinfo {author} {\bibfnamefont {S.}~\bibnamefont {Rogge}},\ and\ \bibinfo {author} {\bibfnamefont {J.~J.}\ \bibnamefont {Longdell}},\ }\bibfield  {title} {\bibinfo {title} {Ultrastrong coupling between a microwave resonator and antiferromagnetic resonances of rare-earth ion spins},\ }\href {https://doi.org/10.1103/PhysRevB.101.214414} {\bibfield  {journal} {\bibinfo  {journal} {Phys. Rev. B}\ }\textbf {\bibinfo {volume} {101}},\ \bibinfo {pages} {214414} (\bibinfo {year} {2020})}\BibitemShut {NoStop}%
\bibitem [{\citenamefont {Potts}\ and\ \citenamefont {Davis}(2020)}]{Potts20}%
  \BibitemOpen
  \bibfield  {author} {\bibinfo {author} {\bibfnamefont {C.~A.}\ \bibnamefont {Potts}}\ and\ \bibinfo {author} {\bibfnamefont {J.~P.}\ \bibnamefont {Davis}},\ }\bibfield  {title} {\bibinfo {title} {Strong magnon–photon coupling within a tunable cryogenic microwave cavity},\ }\href {https://doi.org/10.1063/5.0015660} {\bibfield  {journal} {\bibinfo  {journal} {Applied Physics Letters}\ }\textbf {\bibinfo {volume} {116}},\ \bibinfo {pages} {263503} (\bibinfo {year} {2020})}\BibitemShut {NoStop}%
\bibitem [{\citenamefont {Lachance-Quirion}\ \emph {et~al.}(2020)\citenamefont {Lachance-Quirion}, \citenamefont {Wolski}, \citenamefont {Tabuchi}, \citenamefont {Kono}, \citenamefont {Usami},\ and\ \citenamefont {Nakamura}}]{Lachance-Quirion20}%
  \BibitemOpen
  \bibfield  {author} {\bibinfo {author} {\bibfnamefont {D.}~\bibnamefont {Lachance-Quirion}}, \bibinfo {author} {\bibfnamefont {S.~P.}\ \bibnamefont {Wolski}}, \bibinfo {author} {\bibfnamefont {Y.}~\bibnamefont {Tabuchi}}, \bibinfo {author} {\bibfnamefont {S.}~\bibnamefont {Kono}}, \bibinfo {author} {\bibfnamefont {K.}~\bibnamefont {Usami}},\ and\ \bibinfo {author} {\bibfnamefont {Y.}~\bibnamefont {Nakamura}},\ }\bibfield  {title} {\bibinfo {title} {Entanglement-based single-shot detection of a single magnon with a superconducting qubit},\ }\href {https://doi.org/10.1126/science.aaz9236} {\bibfield  {journal} {\bibinfo  {journal} {Science}\ }\textbf {\bibinfo {volume} {367}},\ \bibinfo {pages} {425} (\bibinfo {year} {2020})}\BibitemShut {NoStop}%
\bibitem [{\citenamefont {Li}\ \emph {et~al.}(2020{\natexlab{a}})\citenamefont {Li}, \citenamefont {Zhang}, \citenamefont {Tyberkevych}, \citenamefont {Kwok}, \citenamefont {Hoffmann},\ and\ \citenamefont {Novosad}}]{Li20JAP}%
  \BibitemOpen
  \bibfield  {author} {\bibinfo {author} {\bibfnamefont {Y.}~\bibnamefont {Li}}, \bibinfo {author} {\bibfnamefont {W.}~\bibnamefont {Zhang}}, \bibinfo {author} {\bibfnamefont {V.}~\bibnamefont {Tyberkevych}}, \bibinfo {author} {\bibfnamefont {W.-K.}\ \bibnamefont {Kwok}}, \bibinfo {author} {\bibfnamefont {A.}~\bibnamefont {Hoffmann}},\ and\ \bibinfo {author} {\bibfnamefont {V.}~\bibnamefont {Novosad}},\ }\bibfield  {title} {\bibinfo {title} {Hybrid magnonics: Physics, circuits, and applications for coherent information processing},\ }\href {https://doi.org/10.1063/5.0020277} {\bibfield  {journal} {\bibinfo  {journal} {Journal of Applied Physics}\ }\textbf {\bibinfo {volume} {128}},\ \bibinfo {pages} {130902} (\bibinfo {year} {2020}{\natexlab{a}})}\BibitemShut {NoStop}%
\bibitem [{\citenamefont {Bhoi}\ \emph {et~al.}(2021)\citenamefont {Bhoi}, \citenamefont {Jang}, \citenamefont {Kim},\ and\ \citenamefont {Kim}}]{Bhoi21}%
  \BibitemOpen
  \bibfield  {author} {\bibinfo {author} {\bibfnamefont {B.}~\bibnamefont {Bhoi}}, \bibinfo {author} {\bibfnamefont {S.-H.}\ \bibnamefont {Jang}}, \bibinfo {author} {\bibfnamefont {B.}~\bibnamefont {Kim}},\ and\ \bibinfo {author} {\bibfnamefont {S.-K.}\ \bibnamefont {Kim}},\ }\bibfield  {title} {\bibinfo {title} {Broadband photon–magnon coupling using arrays of photon resonators},\ }\href {https://doi.org/10.1063/5.0040194} {\bibfield  {journal} {\bibinfo  {journal} {Journal of Applied Physics}\ }\textbf {\bibinfo {volume} {129}},\ \bibinfo {pages} {083904} (\bibinfo {year} {2021})}\BibitemShut {NoStop}%
\bibitem [{\citenamefont {Awschalom}\ and\ \citenamefont {Flatt{\'e}}(2007)}]{Awschalom07}%
  \BibitemOpen
  \bibfield  {author} {\bibinfo {author} {\bibfnamefont {D.~D.}\ \bibnamefont {Awschalom}}\ and\ \bibinfo {author} {\bibfnamefont {M.~E.}\ \bibnamefont {Flatt{\'e}}},\ }\bibfield  {title} {\bibinfo {title} {Challenges for semiconductor spintronics},\ }\href {https://doi.org/10.1038/nphys551} {\bibfield  {journal} {\bibinfo  {journal} {Nature Physics}\ }\textbf {\bibinfo {volume} {3}},\ \bibinfo {pages} {153} (\bibinfo {year} {2007})}\BibitemShut {NoStop}%
\bibitem [{\citenamefont {Kockum}\ \emph {et~al.}(2019)\citenamefont {Kockum}, \citenamefont {Miranowicz}, \citenamefont {De~Liberato}, \citenamefont {Savasta},\ and\ \citenamefont {Nori}}]{Kockum19}%
  \BibitemOpen
  \bibfield  {author} {\bibinfo {author} {\bibfnamefont {A.~F.}\ \bibnamefont {Kockum}}, \bibinfo {author} {\bibfnamefont {A.}~\bibnamefont {Miranowicz}}, \bibinfo {author} {\bibfnamefont {S.}~\bibnamefont {De~Liberato}}, \bibinfo {author} {\bibfnamefont {S.}~\bibnamefont {Savasta}},\ and\ \bibinfo {author} {\bibfnamefont {F.}~\bibnamefont {Nori}},\ }\bibfield  {title} {\bibinfo {title} {Ultrastrong coupling between light and matter},\ }\href {https://doi.org/10.1038/s42254-018-0006-2} {\bibfield  {journal} {\bibinfo  {journal} {Nature Reviews Physics}\ }\textbf {\bibinfo {volume} {1}},\ \bibinfo {pages} {19} (\bibinfo {year} {2019})}\BibitemShut {NoStop}%
\bibitem [{\citenamefont {Roux}\ \emph {et~al.}(2020)\citenamefont {Roux}, \citenamefont {Konishi}, \citenamefont {Helson},\ and\ \citenamefont {Brantut}}]{Roux20}%
  \BibitemOpen
  \bibfield  {author} {\bibinfo {author} {\bibfnamefont {K.}~\bibnamefont {Roux}}, \bibinfo {author} {\bibfnamefont {H.}~\bibnamefont {Konishi}}, \bibinfo {author} {\bibfnamefont {V.}~\bibnamefont {Helson}},\ and\ \bibinfo {author} {\bibfnamefont {J.-P.}\ \bibnamefont {Brantut}},\ }\bibfield  {title} {\bibinfo {title} {Strongly correlated fermions strongly coupled to light},\ }\href {https://doi.org/10.1038/s41467-020-16767-8} {\bibfield  {journal} {\bibinfo  {journal} {Nature Communications}\ }\textbf {\bibinfo {volume} {11}},\ \bibinfo {pages} {2974} (\bibinfo {year} {2020})}\BibitemShut {NoStop}%
\bibitem [{\citenamefont {Yuan}\ \emph {et~al.}(2022)\citenamefont {Yuan}, \citenamefont {Cao}, \citenamefont {Kamra}, \citenamefont {Duine},\ and\ \citenamefont {Yan}}]{Yuan22}%
  \BibitemOpen
  \bibfield  {author} {\bibinfo {author} {\bibfnamefont {H.}~\bibnamefont {Yuan}}, \bibinfo {author} {\bibfnamefont {Y.}~\bibnamefont {Cao}}, \bibinfo {author} {\bibfnamefont {A.}~\bibnamefont {Kamra}}, \bibinfo {author} {\bibfnamefont {R.~A.}\ \bibnamefont {Duine}},\ and\ \bibinfo {author} {\bibfnamefont {P.}~\bibnamefont {Yan}},\ }\bibfield  {title} {\bibinfo {title} {Quantum magnonics: When magnon spintronics meets quantum information science},\ }\href {https://doi.org/https://doi.org/10.1016/j.physrep.2022.03.002} {\bibfield  {journal} {\bibinfo  {journal} {Physics Reports}\ }\textbf {\bibinfo {volume} {965}},\ \bibinfo {pages} {1} (\bibinfo {year} {2022})}\BibitemShut {NoStop}%
\bibitem [{\citenamefont {Spethmann}\ \emph {et~al.}(2016)\citenamefont {Spethmann}, \citenamefont {Kohler}, \citenamefont {Schreppler}, \citenamefont {Buchmann},\ and\ \citenamefont {Stamper-Kurn}}]{Spethmann16}%
  \BibitemOpen
  \bibfield  {author} {\bibinfo {author} {\bibfnamefont {N.}~\bibnamefont {Spethmann}}, \bibinfo {author} {\bibfnamefont {J.}~\bibnamefont {Kohler}}, \bibinfo {author} {\bibfnamefont {S.}~\bibnamefont {Schreppler}}, \bibinfo {author} {\bibfnamefont {L.}~\bibnamefont {Buchmann}},\ and\ \bibinfo {author} {\bibfnamefont {D.~M.}\ \bibnamefont {Stamper-Kurn}},\ }\bibfield  {title} {\bibinfo {title} {Cavity-mediated coupling of mechanical oscillators limited by quantum back-action},\ }\href {https://doi.org/10.1038/nphys3515} {\bibfield  {journal} {\bibinfo  {journal} {Nature Physics}\ }\textbf {\bibinfo {volume} {12}},\ \bibinfo {pages} {27} (\bibinfo {year} {2016})}\BibitemShut {NoStop}%
\bibitem [{\citenamefont {{\'{S}}ciesiek}\ \emph {et~al.}(2020)\citenamefont {{\'{S}}ciesiek}, \citenamefont {Sawicki}, \citenamefont {Pacuski}, \citenamefont {Sobczak}, \citenamefont {Kazimierczuk}, \citenamefont {Golnik},\ and\ \citenamefont {Suffczy{\'{n}}ski}}]{Sciesiek20}%
  \BibitemOpen
  \bibfield  {author} {\bibinfo {author} {\bibfnamefont {M.}~\bibnamefont {{\'{S}}ciesiek}}, \bibinfo {author} {\bibfnamefont {K.}~\bibnamefont {Sawicki}}, \bibinfo {author} {\bibfnamefont {W.}~\bibnamefont {Pacuski}}, \bibinfo {author} {\bibfnamefont {K.}~\bibnamefont {Sobczak}}, \bibinfo {author} {\bibfnamefont {T.}~\bibnamefont {Kazimierczuk}}, \bibinfo {author} {\bibfnamefont {A.}~\bibnamefont {Golnik}},\ and\ \bibinfo {author} {\bibfnamefont {J.}~\bibnamefont {Suffczy{\'{n}}ski}},\ }\bibfield  {title} {\bibinfo {title} {Long-distance coupling and energy transfer between exciton states in magnetically controlled microcavities},\ }\href {https://doi.org/10.1038/s43246-020-00079-x} {\bibfield  {journal} {\bibinfo  {journal} {Communications Materials}\ }\textbf {\bibinfo {volume} {1}},\ \bibinfo {pages} {78} (\bibinfo {year} {2020})}\BibitemShut {NoStop}%
\bibitem [{\citenamefont {Xu}\ \emph {et~al.}(2019)\citenamefont {Xu}, \citenamefont {Rao}, \citenamefont {Gui}, \citenamefont {Jin},\ and\ \citenamefont {Hu}}]{Xu19}%
  \BibitemOpen
  \bibfield  {author} {\bibinfo {author} {\bibfnamefont {P.-C.}\ \bibnamefont {Xu}}, \bibinfo {author} {\bibfnamefont {J.~W.}\ \bibnamefont {Rao}}, \bibinfo {author} {\bibfnamefont {Y.~S.}\ \bibnamefont {Gui}}, \bibinfo {author} {\bibfnamefont {X.}~\bibnamefont {Jin}},\ and\ \bibinfo {author} {\bibfnamefont {C.-M.}\ \bibnamefont {Hu}},\ }\bibfield  {title} {\bibinfo {title} {Cavity-mediated dissipative coupling of distant magnetic moments: Theory and experiment},\ }\href {https://doi.org/10.1103/PhysRevB.100.094415} {\bibfield  {journal} {\bibinfo  {journal} {Phys. Rev. B}\ }\textbf {\bibinfo {volume} {100}},\ \bibinfo {pages} {094415} (\bibinfo {year} {2019})}\BibitemShut {NoStop}%
\bibitem [{\citenamefont {Li}\ \emph {et~al.}(2022)\citenamefont {Li}, \citenamefont {Yefremenko}, \citenamefont {Lisovenko}, \citenamefont {Trevillian}, \citenamefont {Polakovic}, \citenamefont {Cecil}, \citenamefont {Barry}, \citenamefont {Pearson}, \citenamefont {Divan}, \citenamefont {Tyberkevych}, \citenamefont {Chang}, \citenamefont {Welp}, \citenamefont {Kwok},\ and\ \citenamefont {Novosad}}]{Li22}%
  \BibitemOpen
  \bibfield  {author} {\bibinfo {author} {\bibfnamefont {Y.}~\bibnamefont {Li}}, \bibinfo {author} {\bibfnamefont {V.~G.}\ \bibnamefont {Yefremenko}}, \bibinfo {author} {\bibfnamefont {M.}~\bibnamefont {Lisovenko}}, \bibinfo {author} {\bibfnamefont {C.}~\bibnamefont {Trevillian}}, \bibinfo {author} {\bibfnamefont {T.}~\bibnamefont {Polakovic}}, \bibinfo {author} {\bibfnamefont {T.~W.}\ \bibnamefont {Cecil}}, \bibinfo {author} {\bibfnamefont {P.~S.}\ \bibnamefont {Barry}}, \bibinfo {author} {\bibfnamefont {J.}~\bibnamefont {Pearson}}, \bibinfo {author} {\bibfnamefont {R.}~\bibnamefont {Divan}}, \bibinfo {author} {\bibfnamefont {V.}~\bibnamefont {Tyberkevych}}, \bibinfo {author} {\bibfnamefont {C.~L.}\ \bibnamefont {Chang}}, \bibinfo {author} {\bibfnamefont {U.}~\bibnamefont {Welp}}, \bibinfo {author} {\bibfnamefont {W.-K.}\ \bibnamefont {Kwok}},\ and\ \bibinfo {author} {\bibfnamefont {V.}~\bibnamefont {Novosad}},\ }\bibfield  {title} {\bibinfo {title} {Coherent coupling of two remote magnonic resonators
  mediated by superconducting circuits},\ }\href {https://doi.org/10.1103/PhysRevLett.128.047701} {\bibfield  {journal} {\bibinfo  {journal} {Phys. Rev. Lett.}\ }\textbf {\bibinfo {volume} {128}},\ \bibinfo {pages} {047701} (\bibinfo {year} {2022})}\BibitemShut {NoStop}%
\bibitem [{\citenamefont {Wang}\ \emph {et~al.}(2022)\citenamefont {Wang}, \citenamefont {He}, \citenamefont {Yuan}, \citenamefont {Wang}, \citenamefont {Wang}, \citenamefont {Zhang}, \citenamefont {Medlej}, \citenamefont {Chen}, \citenamefont {Yu}, \citenamefont {Han}, \citenamefont {Ansermet},\ and\ \citenamefont {Yu}}]{Hanchen22}%
  \BibitemOpen
  \bibfield  {author} {\bibinfo {author} {\bibfnamefont {H.}~\bibnamefont {Wang}}, \bibinfo {author} {\bibfnamefont {W.}~\bibnamefont {He}}, \bibinfo {author} {\bibfnamefont {R.}~\bibnamefont {Yuan}}, \bibinfo {author} {\bibfnamefont {Y.}~\bibnamefont {Wang}}, \bibinfo {author} {\bibfnamefont {J.}~\bibnamefont {Wang}}, \bibinfo {author} {\bibfnamefont {Y.}~\bibnamefont {Zhang}}, \bibinfo {author} {\bibfnamefont {I.}~\bibnamefont {Medlej}}, \bibinfo {author} {\bibfnamefont {J.}~\bibnamefont {Chen}}, \bibinfo {author} {\bibfnamefont {G.}~\bibnamefont {Yu}}, \bibinfo {author} {\bibfnamefont {X.}~\bibnamefont {Han}}, \bibinfo {author} {\bibfnamefont {J.-P.}\ \bibnamefont {Ansermet}},\ and\ \bibinfo {author} {\bibfnamefont {H.}~\bibnamefont {Yu}},\ }\bibfield  {title} {\bibinfo {title} {Hybridized propagating spin waves in a cofeb/irmn bilayer},\ }\href {https://doi.org/10.1103/PhysRevB.106.064410} {\bibfield  {journal} {\bibinfo  {journal} {Phys. Rev. B}\ }\textbf {\bibinfo {volume} {106}},\ \bibinfo {pages}
  {064410} (\bibinfo {year} {2022})}\BibitemShut {NoStop}%
\bibitem [{\citenamefont {Harvey-Collard}\ \emph {et~al.}(2022)\citenamefont {Harvey-Collard}, \citenamefont {Dijkema}, \citenamefont {Zheng}, \citenamefont {Sammak}, \citenamefont {Scappucci},\ and\ \citenamefont {Vandersypen}}]{Harvey-Collard22}%
  \BibitemOpen
  \bibfield  {author} {\bibinfo {author} {\bibfnamefont {P.}~\bibnamefont {Harvey-Collard}}, \bibinfo {author} {\bibfnamefont {J.}~\bibnamefont {Dijkema}}, \bibinfo {author} {\bibfnamefont {G.}~\bibnamefont {Zheng}}, \bibinfo {author} {\bibfnamefont {A.}~\bibnamefont {Sammak}}, \bibinfo {author} {\bibfnamefont {G.}~\bibnamefont {Scappucci}},\ and\ \bibinfo {author} {\bibfnamefont {L.~M.~K.}\ \bibnamefont {Vandersypen}},\ }\bibfield  {title} {\bibinfo {title} {Coherent spin-spin coupling mediated by virtual microwave photons},\ }\href {https://doi.org/10.1103/PhysRevX.12.021026} {\bibfield  {journal} {\bibinfo  {journal} {Phys. Rev. X}\ }\textbf {\bibinfo {volume} {12}},\ \bibinfo {pages} {021026} (\bibinfo {year} {2022})}\BibitemShut {NoStop}%
\bibitem [{\citenamefont {Nair}\ \emph {et~al.}(2022)\citenamefont {Nair}, \citenamefont {Mukhopadhyay},\ and\ \citenamefont {Agarwal}}]{Nair22}%
  \BibitemOpen
  \bibfield  {author} {\bibinfo {author} {\bibfnamefont {J.~M.~P.}\ \bibnamefont {Nair}}, \bibinfo {author} {\bibfnamefont {D.}~\bibnamefont {Mukhopadhyay}},\ and\ \bibinfo {author} {\bibfnamefont {G.~S.}\ \bibnamefont {Agarwal}},\ }\bibfield  {title} {\bibinfo {title} {Cavity-mediated level attraction and repulsion between magnons},\ }\href {https://doi.org/10.1103/PhysRevB.105.214418} {\bibfield  {journal} {\bibinfo  {journal} {Phys. Rev. B}\ }\textbf {\bibinfo {volume} {105}},\ \bibinfo {pages} {214418} (\bibinfo {year} {2022})}\BibitemShut {NoStop}%
\bibitem [{\citenamefont {Yang}\ \emph {et~al.}(2022)\citenamefont {Yang}, \citenamefont {Liu},\ and\ \citenamefont {Yang}}]{Yang22}%
  \BibitemOpen
  \bibfield  {author} {\bibinfo {author} {\bibfnamefont {Z.-B.}\ \bibnamefont {Yang}}, \bibinfo {author} {\bibfnamefont {H.-Y.}\ \bibnamefont {Liu}},\ and\ \bibinfo {author} {\bibfnamefont {R.-C.}\ \bibnamefont {Yang}},\ }\bibfield  {title} {\bibinfo {title} {Asymmetric quantum synchronization generation in antiferromagnet-cavity systems},\ }\href {https://doi.org/10.1140/epjp/s13360-022-03064-0} {\bibfield  {journal} {\bibinfo  {journal} {The European Physical Journal Plus}\ }\textbf {\bibinfo {volume} {137}},\ \bibinfo {pages} {878} (\bibinfo {year} {2022})}\BibitemShut {NoStop}%
\bibitem [{\citenamefont {Bia\l{}ek}\ \emph {et~al.}(2023)\citenamefont {Bia\l{}ek}, \citenamefont {Knap},\ and\ \citenamefont {Ansermet}}]{Bialek23}%
  \BibitemOpen
  \bibfield  {author} {\bibinfo {author} {\bibfnamefont {M.}~\bibnamefont {Bia\l{}ek}}, \bibinfo {author} {\bibfnamefont {W.}~\bibnamefont {Knap}},\ and\ \bibinfo {author} {\bibfnamefont {J.-P.}\ \bibnamefont {Ansermet}},\ }\bibfield  {title} {\bibinfo {title} {Cavity-mediated coupling of terahertz antiferromagnetic resonators},\ }\href {https://doi.org/10.1103/PhysRevApplied.19.064007} {\bibfield  {journal} {\bibinfo  {journal} {Phys. Rev. Appl.}\ }\textbf {\bibinfo {volume} {19}},\ \bibinfo {pages} {064007} (\bibinfo {year} {2023})}\BibitemShut {NoStop}%
\bibitem [{\citenamefont {Berk}\ \emph {et~al.}(2019)\citenamefont {Berk}, \citenamefont {Jaris}, \citenamefont {Yang}, \citenamefont {Dhuey}, \citenamefont {Cabrini},\ and\ \citenamefont {Schmidt}}]{Berk19}%
  \BibitemOpen
  \bibfield  {author} {\bibinfo {author} {\bibfnamefont {C.}~\bibnamefont {Berk}}, \bibinfo {author} {\bibfnamefont {M.}~\bibnamefont {Jaris}}, \bibinfo {author} {\bibfnamefont {W.}~\bibnamefont {Yang}}, \bibinfo {author} {\bibfnamefont {S.}~\bibnamefont {Dhuey}}, \bibinfo {author} {\bibfnamefont {S.}~\bibnamefont {Cabrini}},\ and\ \bibinfo {author} {\bibfnamefont {H.}~\bibnamefont {Schmidt}},\ }\bibfield  {title} {\bibinfo {title} {Strongly coupled magnon--phonon dynamics in a single nanomagnet},\ }\href {https://doi.org/10.1038/s41467-019-10545-x} {\bibfield  {journal} {\bibinfo  {journal} {Nature Communications}\ }\textbf {\bibinfo {volume} {10}},\ \bibinfo {pages} {2652} (\bibinfo {year} {2019})}\BibitemShut {NoStop}%
\bibitem [{\citenamefont {Li}\ \emph {et~al.}(2021)\citenamefont {Li}, \citenamefont {Zhao}, \citenamefont {Zhang}, \citenamefont {Hoffmann},\ and\ \citenamefont {Novosad}}]{Li21}%
  \BibitemOpen
  \bibfield  {author} {\bibinfo {author} {\bibfnamefont {Y.}~\bibnamefont {Li}}, \bibinfo {author} {\bibfnamefont {C.}~\bibnamefont {Zhao}}, \bibinfo {author} {\bibfnamefont {W.}~\bibnamefont {Zhang}}, \bibinfo {author} {\bibfnamefont {A.}~\bibnamefont {Hoffmann}},\ and\ \bibinfo {author} {\bibfnamefont {V.}~\bibnamefont {Novosad}},\ }\bibfield  {title} {\bibinfo {title} {{Advances in coherent coupling between magnons and acoustic phonons}},\ }\href {https://doi.org/10.1063/5.0047054} {\bibfield  {journal} {\bibinfo  {journal} {APL Materials}\ }\textbf {\bibinfo {volume} {9}},\ \bibinfo {pages} {060902} (\bibinfo {year} {2021})},\ \Eprint {https://arxiv.org/abs/https://pubs.aip.org/aip/apm/article-pdf/doi/10.1063/5.0047054/14565679/060902\_1\_online.pdf} {https://pubs.aip.org/aip/apm/article-pdf/doi/10.1063/5.0047054/14565679/060902\_1\_online.pdf} \BibitemShut {NoStop}%
\bibitem [{\citenamefont {Khan}\ \emph {et~al.}(2020)\citenamefont {Khan}, \citenamefont {Kanamaru}, \citenamefont {Matsumoto}, \citenamefont {Ito},\ and\ \citenamefont {Satoh}}]{Khan20}%
  \BibitemOpen
  \bibfield  {author} {\bibinfo {author} {\bibfnamefont {P.}~\bibnamefont {Khan}}, \bibinfo {author} {\bibfnamefont {M.}~\bibnamefont {Kanamaru}}, \bibinfo {author} {\bibfnamefont {K.}~\bibnamefont {Matsumoto}}, \bibinfo {author} {\bibfnamefont {T.}~\bibnamefont {Ito}},\ and\ \bibinfo {author} {\bibfnamefont {T.}~\bibnamefont {Satoh}},\ }\bibfield  {title} {\bibinfo {title} {Ultrafast light-driven simultaneous excitation of coherent terahertz magnons and phonons in multiferroic $\mathrm{BiFe}{\mathrm{o}}_{3}$},\ }\href {https://doi.org/10.1103/PhysRevB.101.134413} {\bibfield  {journal} {\bibinfo  {journal} {Phys. Rev. B}\ }\textbf {\bibinfo {volume} {101}},\ \bibinfo {pages} {134413} (\bibinfo {year} {2020})}\BibitemShut {NoStop}%
\bibitem [{\citenamefont {Li}\ \emph {et~al.}(2020{\natexlab{b}})\citenamefont {Li}, \citenamefont {Simensen}, \citenamefont {Reitz}, \citenamefont {Sun}, \citenamefont {Yuan}, \citenamefont {Li}, \citenamefont {Tserkovnyak}, \citenamefont {Brataas},\ and\ \citenamefont {Shi}}]{Li20PRL}%
  \BibitemOpen
  \bibfield  {author} {\bibinfo {author} {\bibfnamefont {J.}~\bibnamefont {Li}}, \bibinfo {author} {\bibfnamefont {H.~T.}\ \bibnamefont {Simensen}}, \bibinfo {author} {\bibfnamefont {D.}~\bibnamefont {Reitz}}, \bibinfo {author} {\bibfnamefont {Q.}~\bibnamefont {Sun}}, \bibinfo {author} {\bibfnamefont {W.}~\bibnamefont {Yuan}}, \bibinfo {author} {\bibfnamefont {C.}~\bibnamefont {Li}}, \bibinfo {author} {\bibfnamefont {Y.}~\bibnamefont {Tserkovnyak}}, \bibinfo {author} {\bibfnamefont {A.}~\bibnamefont {Brataas}},\ and\ \bibinfo {author} {\bibfnamefont {J.}~\bibnamefont {Shi}},\ }\bibfield  {title} {\bibinfo {title} {Observation of magnon polarons in a uniaxial antiferromagnetic insulator},\ }\href {https://doi.org/10.1103/PhysRevLett.125.217201} {\bibfield  {journal} {\bibinfo  {journal} {Phys. Rev. Lett.}\ }\textbf {\bibinfo {volume} {125}},\ \bibinfo {pages} {217201} (\bibinfo {year} {2020}{\natexlab{b}})}\BibitemShut {NoStop}%
\bibitem [{\citenamefont {Liu}\ \emph {et~al.}(2021)\citenamefont {Liu}, \citenamefont {Granados~del \'Aguila}, \citenamefont {Bhowmick}, \citenamefont {Gan}, \citenamefont {Thu Ha~Do}, \citenamefont {Prosnikov}, \citenamefont {Sedmidubsk\'y}, \citenamefont {Sofer}, \citenamefont {Christianen}, \citenamefont {Sengupta},\ and\ \citenamefont {Xiong}}]{Liu21}%
  \BibitemOpen
  \bibfield  {author} {\bibinfo {author} {\bibfnamefont {S.}~\bibnamefont {Liu}}, \bibinfo {author} {\bibfnamefont {A.}~\bibnamefont {Granados~del \'Aguila}}, \bibinfo {author} {\bibfnamefont {D.}~\bibnamefont {Bhowmick}}, \bibinfo {author} {\bibfnamefont {C.~K.}\ \bibnamefont {Gan}}, \bibinfo {author} {\bibfnamefont {T.}~\bibnamefont {Thu Ha~Do}}, \bibinfo {author} {\bibfnamefont {M.~A.}\ \bibnamefont {Prosnikov}}, \bibinfo {author} {\bibfnamefont {D.}~\bibnamefont {Sedmidubsk\'y}}, \bibinfo {author} {\bibfnamefont {Z.}~\bibnamefont {Sofer}}, \bibinfo {author} {\bibfnamefont {P.~C.~M.}\ \bibnamefont {Christianen}}, \bibinfo {author} {\bibfnamefont {P.}~\bibnamefont {Sengupta}},\ and\ \bibinfo {author} {\bibfnamefont {Q.}~\bibnamefont {Xiong}},\ }\bibfield  {title} {\bibinfo {title} {Direct observation of magnon-phonon strong coupling in two-dimensional antiferromagnet at high magnetic fields},\ }\href {https://doi.org/10.1103/PhysRevLett.127.097401} {\bibfield  {journal} {\bibinfo  {journal} {Phys. Rev.
  Lett.}\ }\textbf {\bibinfo {volume} {127}},\ \bibinfo {pages} {097401} (\bibinfo {year} {2021})}\BibitemShut {NoStop}%
\bibitem [{\citenamefont {Vaclavkova}\ \emph {et~al.}(2021)\citenamefont {Vaclavkova}, \citenamefont {Palit}, \citenamefont {Wyzula}, \citenamefont {Ghosh}, \citenamefont {Delhomme}, \citenamefont {Maity}, \citenamefont {Kapuscinski}, \citenamefont {Ghosh}, \citenamefont {Veis}, \citenamefont {Grzeszczyk}, \citenamefont {Faugeras}, \citenamefont {Orlita}, \citenamefont {Datta},\ and\ \citenamefont {Potemski}}]{Vaclakova21}%
  \BibitemOpen
  \bibfield  {author} {\bibinfo {author} {\bibfnamefont {D.}~\bibnamefont {Vaclavkova}}, \bibinfo {author} {\bibfnamefont {M.}~\bibnamefont {Palit}}, \bibinfo {author} {\bibfnamefont {J.}~\bibnamefont {Wyzula}}, \bibinfo {author} {\bibfnamefont {S.}~\bibnamefont {Ghosh}}, \bibinfo {author} {\bibfnamefont {A.}~\bibnamefont {Delhomme}}, \bibinfo {author} {\bibfnamefont {S.}~\bibnamefont {Maity}}, \bibinfo {author} {\bibfnamefont {P.}~\bibnamefont {Kapuscinski}}, \bibinfo {author} {\bibfnamefont {A.}~\bibnamefont {Ghosh}}, \bibinfo {author} {\bibfnamefont {M.}~\bibnamefont {Veis}}, \bibinfo {author} {\bibfnamefont {M.}~\bibnamefont {Grzeszczyk}}, \bibinfo {author} {\bibfnamefont {C.}~\bibnamefont {Faugeras}}, \bibinfo {author} {\bibfnamefont {M.}~\bibnamefont {Orlita}}, \bibinfo {author} {\bibfnamefont {S.}~\bibnamefont {Datta}},\ and\ \bibinfo {author} {\bibfnamefont {M.}~\bibnamefont {Potemski}},\ }\bibfield  {title} {\bibinfo {title} {Magnon polarons in the van der waals antiferromagnet
  $\mathrm{Fe}{\mathrm{ps}}_{3}$},\ }\href {https://doi.org/10.1103/PhysRevB.104.134437} {\bibfield  {journal} {\bibinfo  {journal} {Phys. Rev. B}\ }\textbf {\bibinfo {volume} {104}},\ \bibinfo {pages} {134437} (\bibinfo {year} {2021})}\BibitemShut {NoStop}%
\bibitem [{\citenamefont {Diederich}\ \emph {et~al.}(2022)\citenamefont {Diederich}, \citenamefont {Cenker}, \citenamefont {Ren}, \citenamefont {Fonseca}, \citenamefont {Chica}, \citenamefont {Bae}, \citenamefont {Zhu}, \citenamefont {Roy}, \citenamefont {Cao}, \citenamefont {Xiao},\ and\ \citenamefont {Xu}}]{Diederich22}%
  \BibitemOpen
  \bibfield  {author} {\bibinfo {author} {\bibfnamefont {G.~M.}\ \bibnamefont {Diederich}}, \bibinfo {author} {\bibfnamefont {J.}~\bibnamefont {Cenker}}, \bibinfo {author} {\bibfnamefont {Y.}~\bibnamefont {Ren}}, \bibinfo {author} {\bibfnamefont {J.}~\bibnamefont {Fonseca}}, \bibinfo {author} {\bibfnamefont {D.~G.}\ \bibnamefont {Chica}}, \bibinfo {author} {\bibfnamefont {Y.~J.}\ \bibnamefont {Bae}}, \bibinfo {author} {\bibfnamefont {X.}~\bibnamefont {Zhu}}, \bibinfo {author} {\bibfnamefont {X.}~\bibnamefont {Roy}}, \bibinfo {author} {\bibfnamefont {T.}~\bibnamefont {Cao}}, \bibinfo {author} {\bibfnamefont {D.}~\bibnamefont {Xiao}},\ and\ \bibinfo {author} {\bibfnamefont {X.}~\bibnamefont {Xu}},\ }\bibfield  {title} {\bibinfo {title} {Tunable interaction between excitons and hybridized magnons in a layered semiconductor},\ }\bibfield  {journal} {\bibinfo  {journal} {Nature Nanotechnology}\ }\href {https://doi.org/10.1038/s41565-022-01259-1} {10.1038/s41565-022-01259-1} (\bibinfo {year} {2022})\BibitemShut
  {NoStop}%
\bibitem [{\citenamefont {Sivarajah}\ \emph {et~al.}(2019)\citenamefont {Sivarajah}, \citenamefont {Steinbacher}, \citenamefont {Dastrup}, \citenamefont {Lu}, \citenamefont {Xiang}, \citenamefont {Ren}, \citenamefont {Kamba}, \citenamefont {Cao},\ and\ \citenamefont {Nelson}}]{Sivarajah19}%
  \BibitemOpen
  \bibfield  {author} {\bibinfo {author} {\bibfnamefont {P.}~\bibnamefont {Sivarajah}}, \bibinfo {author} {\bibfnamefont {A.}~\bibnamefont {Steinbacher}}, \bibinfo {author} {\bibfnamefont {B.}~\bibnamefont {Dastrup}}, \bibinfo {author} {\bibfnamefont {J.}~\bibnamefont {Lu}}, \bibinfo {author} {\bibfnamefont {M.}~\bibnamefont {Xiang}}, \bibinfo {author} {\bibfnamefont {W.}~\bibnamefont {Ren}}, \bibinfo {author} {\bibfnamefont {S.}~\bibnamefont {Kamba}}, \bibinfo {author} {\bibfnamefont {S.}~\bibnamefont {Cao}},\ and\ \bibinfo {author} {\bibfnamefont {K.~A.}\ \bibnamefont {Nelson}},\ }\bibfield  {title} {\bibinfo {title} {{THz-frequency magnon-phonon-polaritons in the collective strong-coupling regime}},\ }\href {https://doi.org/10.1063/1.5083849} {\bibfield  {journal} {\bibinfo  {journal} {Journal of Applied Physics}\ }\textbf {\bibinfo {volume} {125}},\ \bibinfo {pages} {213103} (\bibinfo {year} {2019})}\BibitemShut {NoStop}%
\bibitem [{\citenamefont {Marsh}\ \emph {et~al.}(2019)\citenamefont {Marsh}, \citenamefont {Fong}, \citenamefont {Lentz}, \citenamefont {\ifmmode~\check{S}\else \v{S}\fi{}mejkal},\ and\ \citenamefont {Ali}}]{Marsh19}%
  \BibitemOpen
  \bibfield  {author} {\bibinfo {author} {\bibfnamefont {D.~J.~E.}\ \bibnamefont {Marsh}}, \bibinfo {author} {\bibfnamefont {K.~C.}\ \bibnamefont {Fong}}, \bibinfo {author} {\bibfnamefont {E.~W.}\ \bibnamefont {Lentz}}, \bibinfo {author} {\bibfnamefont {L.}~\bibnamefont {\ifmmode~\check{S}\else \v{S}\fi{}mejkal}},\ and\ \bibinfo {author} {\bibfnamefont {M.~N.}\ \bibnamefont {Ali}},\ }\bibfield  {title} {\bibinfo {title} {Proposal to detect dark matter using axionic topological antiferromagnets},\ }\href {https://doi.org/10.1103/PhysRevLett.123.121601} {\bibfield  {journal} {\bibinfo  {journal} {Phys. Rev. Lett.}\ }\textbf {\bibinfo {volume} {123}},\ \bibinfo {pages} {121601} (\bibinfo {year} {2019})}\BibitemShut {NoStop}%
\bibitem [{\citenamefont {Mitridate}\ \emph {et~al.}(2020)\citenamefont {Mitridate}, \citenamefont {Trickle}, \citenamefont {Zhang},\ and\ \citenamefont {Zurek}}]{Mitridate20}%
  \BibitemOpen
  \bibfield  {author} {\bibinfo {author} {\bibfnamefont {A.}~\bibnamefont {Mitridate}}, \bibinfo {author} {\bibfnamefont {T.}~\bibnamefont {Trickle}}, \bibinfo {author} {\bibfnamefont {Z.}~\bibnamefont {Zhang}},\ and\ \bibinfo {author} {\bibfnamefont {K.~M.}\ \bibnamefont {Zurek}},\ }\bibfield  {title} {\bibinfo {title} {Detectability of axion dark matter with phonon polaritons and magnons},\ }\href {https://doi.org/10.1103/PhysRevD.102.095005} {\bibfield  {journal} {\bibinfo  {journal} {Phys. Rev. D}\ }\textbf {\bibinfo {volume} {102}},\ \bibinfo {pages} {095005} (\bibinfo {year} {2020})}\BibitemShut {NoStop}%
\bibitem [{\citenamefont {Wang}\ \emph {et~al.}(2011)\citenamefont {Wang}, \citenamefont {Li}, \citenamefont {Zhang},\ and\ \citenamefont {Qi}}]{Wang11}%
  \BibitemOpen
  \bibfield  {author} {\bibinfo {author} {\bibfnamefont {J.}~\bibnamefont {Wang}}, \bibinfo {author} {\bibfnamefont {R.}~\bibnamefont {Li}}, \bibinfo {author} {\bibfnamefont {S.-C.}\ \bibnamefont {Zhang}},\ and\ \bibinfo {author} {\bibfnamefont {X.-L.}\ \bibnamefont {Qi}},\ }\bibfield  {title} {\bibinfo {title} {Topological magnetic insulators with corundum structure},\ }\href {https://doi.org/10.1103/PhysRevLett.106.126403} {\bibfield  {journal} {\bibinfo  {journal} {Phys. Rev. Lett.}\ }\textbf {\bibinfo {volume} {106}},\ \bibinfo {pages} {126403} (\bibinfo {year} {2011})}\BibitemShut {NoStop}%
\bibitem [{\citenamefont {Karaki}\ \emph {et~al.}(2022)\citenamefont {Karaki}, \citenamefont {Yang}, \citenamefont {Williams}, \citenamefont {Nawwar}, \citenamefont {Doan-Nguyen}, \citenamefont {Goldberger},\ and\ \citenamefont {Lu}}]{Karaki22}%
  \BibitemOpen
  \bibfield  {author} {\bibinfo {author} {\bibfnamefont {M.~J.}\ \bibnamefont {Karaki}}, \bibinfo {author} {\bibfnamefont {X.}~\bibnamefont {Yang}}, \bibinfo {author} {\bibfnamefont {A.~J.}\ \bibnamefont {Williams}}, \bibinfo {author} {\bibfnamefont {M.}~\bibnamefont {Nawwar}}, \bibinfo {author} {\bibfnamefont {V.}~\bibnamefont {Doan-Nguyen}}, \bibinfo {author} {\bibfnamefont {J.~E.}\ \bibnamefont {Goldberger}},\ and\ \bibinfo {author} {\bibfnamefont {Y.-M.}\ \bibnamefont {Lu}},\ }\href {https://doi.org/10.48550/ARXIV.2206.06248} {\bibinfo {title} {An efficient material search for room temperature topological magnons}} (\bibinfo {year} {2022})\BibitemShut {NoStop}%
\bibitem [{\citenamefont {Marsh}\ \emph {et~al.}(2022)\citenamefont {Marsh}, \citenamefont {McDonald}, \citenamefont {Millar},\ and\ \citenamefont {Schütte-Engel}}]{Marsh22}%
  \BibitemOpen
  \bibfield  {author} {\bibinfo {author} {\bibfnamefont {D.~J.~E.}\ \bibnamefont {Marsh}}, \bibinfo {author} {\bibfnamefont {J.~I.}\ \bibnamefont {McDonald}}, \bibinfo {author} {\bibfnamefont {A.~J.}\ \bibnamefont {Millar}},\ and\ \bibinfo {author} {\bibfnamefont {J.}~\bibnamefont {Schütte-Engel}},\ }\href {https://doi.org/10.48550/ARXIV.2209.12909} {\bibinfo {title} {Axion detection with phonon-polaritons revisited}} (\bibinfo {year} {2022})\BibitemShut {NoStop}%
\bibitem [{\citenamefont {Rezende}\ \emph {et~al.}(2019)\citenamefont {Rezende}, \citenamefont {Azevedo},\ and\ \citenamefont {Rodríguez-Suárez}}]{Rezende19}%
  \BibitemOpen
  \bibfield  {author} {\bibinfo {author} {\bibfnamefont {S.~M.}\ \bibnamefont {Rezende}}, \bibinfo {author} {\bibfnamefont {A.}~\bibnamefont {Azevedo}},\ and\ \bibinfo {author} {\bibfnamefont {R.~L.}\ \bibnamefont {Rodríguez-Suárez}},\ }\bibfield  {title} {\bibinfo {title} {{Introduction to antiferromagnetic magnons}},\ }\href {https://doi.org/10.1063/1.5109132} {\bibfield  {journal} {\bibinfo  {journal} {Journal of Applied Physics}\ }\textbf {\bibinfo {volume} {126}},\ \bibinfo {pages} {151101} (\bibinfo {year} {2019})},\ \Eprint {https://arxiv.org/abs/https://pubs.aip.org/aip/jap/article-pdf/doi/10.1063/1.5109132/13019547/151101\_1\_online.pdf} {https://pubs.aip.org/aip/jap/article-pdf/doi/10.1063/1.5109132/13019547/151101\_1\_online.pdf} \BibitemShut {NoStop}%
\bibitem [{\citenamefont {Szwagierczak}\ \emph {et~al.}(2021)\citenamefont {Szwagierczak}, \citenamefont {Synkiewicz-Musialska}, \citenamefont {Kulawik},\ and\ \citenamefont {Pałka}}]{Szwagierczak21}%
  \BibitemOpen
  \bibfield  {author} {\bibinfo {author} {\bibfnamefont {D.}~\bibnamefont {Szwagierczak}}, \bibinfo {author} {\bibfnamefont {B.}~\bibnamefont {Synkiewicz-Musialska}}, \bibinfo {author} {\bibfnamefont {J.}~\bibnamefont {Kulawik}},\ and\ \bibinfo {author} {\bibfnamefont {N.}~\bibnamefont {Pałka}},\ }\bibfield  {title} {\bibinfo {title} {Sintering, microstructure, and dielectric properties of copper borates for high frequency ltcc applications},\ }\bibfield  {journal} {\bibinfo  {journal} {Materials}\ }\textbf {\bibinfo {volume} {14}},\ \href {https://doi.org/10.3390/ma14144017} {10.3390/ma14144017} (\bibinfo {year} {2021})\BibitemShut {NoStop}%
\bibitem [{\citenamefont {Thomas}\ \emph {et~al.}(2020)\citenamefont {Thomas}, \citenamefont {Tan}, \citenamefont {Fernandez},\ and\ \citenamefont {Barnes}}]{Thomas20}%
  \BibitemOpen
  \bibfield  {author} {\bibinfo {author} {\bibfnamefont {P.~A.}\ \bibnamefont {Thomas}}, \bibinfo {author} {\bibfnamefont {W.~J.}\ \bibnamefont {Tan}}, \bibinfo {author} {\bibfnamefont {H.~A.}\ \bibnamefont {Fernandez}},\ and\ \bibinfo {author} {\bibfnamefont {W.~L.}\ \bibnamefont {Barnes}},\ }\bibfield  {title} {\bibinfo {title} {A new signature for strong light--matter coupling using spectroscopic ellipsometry},\ }\href {https://doi.org/10.1021/acs.nanolett.0c01963} {\bibfield  {journal} {\bibinfo  {journal} {Nano Letters}\ }\textbf {\bibinfo {volume} {20}},\ \bibinfo {pages} {6412} (\bibinfo {year} {2020})}\BibitemShut {NoStop}%
\bibitem [{\citenamefont {Roth}(1958)}]{Roth58}%
  \BibitemOpen
  \bibfield  {author} {\bibinfo {author} {\bibfnamefont {W.~L.}\ \bibnamefont {Roth}},\ }\bibfield  {title} {\bibinfo {title} {Magnetic structures of mno, feo, coo, and nio},\ }\href {https://doi.org/10.1103/PhysRev.110.1333} {\bibfield  {journal} {\bibinfo  {journal} {Phys. Rev.}\ }\textbf {\bibinfo {volume} {110}},\ \bibinfo {pages} {1333} (\bibinfo {year} {1958})}\BibitemShut {NoStop}%
\bibitem [{\citenamefont {Massarotti}\ \emph {et~al.}(1991)\citenamefont {Massarotti}, \citenamefont {Capsoni}, \citenamefont {Berbenni}, \citenamefont {Riccardi}, \citenamefont {Marini},\ and\ \citenamefont {Antolini}}]{Massarotti91}%
  \BibitemOpen
  \bibfield  {author} {\bibinfo {author} {\bibfnamefont {V.}~\bibnamefont {Massarotti}}, \bibinfo {author} {\bibfnamefont {D.}~\bibnamefont {Capsoni}}, \bibinfo {author} {\bibfnamefont {V.}~\bibnamefont {Berbenni}}, \bibinfo {author} {\bibfnamefont {R.}~\bibnamefont {Riccardi}}, \bibinfo {author} {\bibfnamefont {A.}~\bibnamefont {Marini}},\ and\ \bibinfo {author} {\bibfnamefont {E.}~\bibnamefont {Antolini}},\ }\bibfield  {title} {\bibinfo {title} {Structural characterization of nickel oxide},\ }\href {https://doi.org/doi:10.1515/zna-1991-0606} {\bibfield  {journal} {\bibinfo  {journal} {Zeitschrift für Naturforschung A}\ }\textbf {\bibinfo {volume} {46}},\ \bibinfo {pages} {503} (\bibinfo {year} {1991})}\BibitemShut {NoStop}%
\bibitem [{\citenamefont {Born}\ and\ \citenamefont {Wolf}(1980)}]{BORN}%
  \BibitemOpen
  \bibfield  {author} {\bibinfo {author} {\bibfnamefont {M.}~\bibnamefont {Born}}\ and\ \bibinfo {author} {\bibfnamefont {E.}~\bibnamefont {Wolf}},\ }\bibfield  {title} {\bibinfo {title} {Chapter 1 - basic properties of the electromagnetic field},\ }in\ \href {https://doi.org/https://doi.org/10.1016/B978-0-08-026482-0.50008-6} {\emph {\bibinfo {booktitle} {Principles of Optics (Sixth Edition)}}},\ \bibinfo {editor} {edited by\ \bibinfo {editor} {\bibfnamefont {M.}~\bibnamefont {Born}}\ and\ \bibinfo {editor} {\bibfnamefont {E.}~\bibnamefont {Wolf}}}\ (\bibinfo  {publisher} {Pergamon},\ \bibinfo {year} {1980})\ \bibinfo {edition} {sixth edition}\ ed.,\ pp.\ \bibinfo {pages} {1 -- 70}\BibitemShut {NoStop}%
\bibitem [{\citenamefont {Todorov}(2015)}]{Todorov_PhysRevB125409_2015}%
  \BibitemOpen
  \bibfield  {author} {\bibinfo {author} {\bibfnamefont {Y.}~\bibnamefont {Todorov}},\ }\bibfield  {title} {\bibinfo {title} {Dipolar quantum electrodynamics of the two-dimensional electron gas},\ }\href {https://doi.org/10.1103/PhysRevB.91.125409} {\bibfield  {journal} {\bibinfo  {journal} {Phys. Rev. B}\ }\textbf {\bibinfo {volume} {91}},\ \bibinfo {pages} {125409} (\bibinfo {year} {2015})}\BibitemShut {NoStop}%
\bibitem [{\citenamefont {Ghosh}\ \emph {et~al.}(2021)\citenamefont {Ghosh}, \citenamefont {Palit}, \citenamefont {Maity}, \citenamefont {Dwij}, \citenamefont {Rana},\ and\ \citenamefont {Datta}}]{Ghosh21}%
  \BibitemOpen
  \bibfield  {author} {\bibinfo {author} {\bibfnamefont {A.}~\bibnamefont {Ghosh}}, \bibinfo {author} {\bibfnamefont {M.}~\bibnamefont {Palit}}, \bibinfo {author} {\bibfnamefont {S.}~\bibnamefont {Maity}}, \bibinfo {author} {\bibfnamefont {V.}~\bibnamefont {Dwij}}, \bibinfo {author} {\bibfnamefont {S.}~\bibnamefont {Rana}},\ and\ \bibinfo {author} {\bibfnamefont {S.}~\bibnamefont {Datta}},\ }\bibfield  {title} {\bibinfo {title} {{Spin-phonon coupling and magnon scattering in few-layer antiferromagnetic ${\mathrm{FePS}}_{3}$}},\ }\href {https://doi.org/10.1103/PhysRevB.103.064431} {\bibfield  {journal} {\bibinfo  {journal} {Phys. Rev. B}\ }\textbf {\bibinfo {volume} {103}},\ \bibinfo {pages} {064431} (\bibinfo {year} {2021})}\BibitemShut {NoStop}%
\bibitem [{\citenamefont {Todorov}\ \emph {et~al.}(2010)\citenamefont {Todorov}, \citenamefont {Andrews}, \citenamefont {Colombelli}, \citenamefont {De~Liberato}, \citenamefont {Ciuti}, \citenamefont {Klang}, \citenamefont {Strasser},\ and\ \citenamefont {Sirtori}}]{Todorov_PRL2010}%
  \BibitemOpen
  \bibfield  {author} {\bibinfo {author} {\bibfnamefont {Y.}~\bibnamefont {Todorov}}, \bibinfo {author} {\bibfnamefont {A.~M.}\ \bibnamefont {Andrews}}, \bibinfo {author} {\bibfnamefont {R.}~\bibnamefont {Colombelli}}, \bibinfo {author} {\bibfnamefont {S.}~\bibnamefont {De~Liberato}}, \bibinfo {author} {\bibfnamefont {C.}~\bibnamefont {Ciuti}}, \bibinfo {author} {\bibfnamefont {P.}~\bibnamefont {Klang}}, \bibinfo {author} {\bibfnamefont {G.}~\bibnamefont {Strasser}},\ and\ \bibinfo {author} {\bibfnamefont {C.}~\bibnamefont {Sirtori}},\ }\bibfield  {title} {\bibinfo {title} {Ultrastrong light-matter coupling regime with polariton dots},\ }\href {https://doi.org/10.1103/PhysRevLett.105.196402} {\bibfield  {journal} {\bibinfo  {journal} {Phys. Rev. Lett.}\ }\textbf {\bibinfo {volume} {\textbf{105}}},\ \bibinfo {pages} {196402} (\bibinfo {year} {(2010)})}\BibitemShut {NoStop}%
\bibitem [{\citenamefont {Jeannin}\ \emph {et~al.}(2019)\citenamefont {Jeannin}, \citenamefont {Mariotti~Nesurini}, \citenamefont {Suffit}, \citenamefont {Gacemi}, \citenamefont {Vasanelli}, \citenamefont {Li}, \citenamefont {Davies}, \citenamefont {Linfield}, \citenamefont {Sirtori},\ and\ \citenamefont {Todorov}}]{Jeannin_2019}%
  \BibitemOpen
  \bibfield  {author} {\bibinfo {author} {\bibfnamefont {M.}~\bibnamefont {Jeannin}}, \bibinfo {author} {\bibfnamefont {G.}~\bibnamefont {Mariotti~Nesurini}}, \bibinfo {author} {\bibfnamefont {S.}~\bibnamefont {Suffit}}, \bibinfo {author} {\bibfnamefont {D.}~\bibnamefont {Gacemi}}, \bibinfo {author} {\bibfnamefont {A.}~\bibnamefont {Vasanelli}}, \bibinfo {author} {\bibfnamefont {L.}~\bibnamefont {Li}}, \bibinfo {author} {\bibfnamefont {A.~G.}\ \bibnamefont {Davies}}, \bibinfo {author} {\bibfnamefont {E.}~\bibnamefont {Linfield}}, \bibinfo {author} {\bibfnamefont {C.}~\bibnamefont {Sirtori}},\ and\ \bibinfo {author} {\bibfnamefont {Y.}~\bibnamefont {Todorov}},\ }\bibfield  {title} {\bibinfo {title} {Ultrastrong light–matter coupling in deeply subwavelength thz lc resonators},\ }\href {https://doi.org/10.1021/acsphotonics.8b01778} {\bibfield  {journal} {\bibinfo  {journal} {ACS Photonics}\ }\textbf {\bibinfo {volume} {6}},\ \bibinfo {pages} {1207} (\bibinfo {year} {2019})},\ \Eprint
  {https://arxiv.org/abs/https://doi.org/10.1021/acsphotonics.8b01778} {https://doi.org/10.1021/acsphotonics.8b01778} \BibitemShut {NoStop}%
\bibitem [{\citenamefont {Sanchez-Manzano}\ \emph {et~al.}(2022)\citenamefont {Sanchez-Manzano}, \citenamefont {Mesoraca}, \citenamefont {Cuellar}, \citenamefont {Cabero}, \citenamefont {Rouco}, \citenamefont {Orfila}, \citenamefont {Palermo}, \citenamefont {Balan}, \citenamefont {Marcano}, \citenamefont {Sander}, \citenamefont {Rocci}, \citenamefont {Garcia-Barriocanal}, \citenamefont {Gallego}, \citenamefont {Tornos}, \citenamefont {Rivera}, \citenamefont {Mompean}, \citenamefont {Garcia-Hernandez}, \citenamefont {Gonzalez-Calbet}, \citenamefont {Leon}, \citenamefont {Valencia}, \citenamefont {Feuillet-Palma}, \citenamefont {Bergeal}, \citenamefont {Buzdin}, \citenamefont {Lesueur}, \citenamefont {Villegas},\ and\ \citenamefont {Santamaria}}]{Sanchez-Manzano2022}%
  \BibitemOpen
  \bibfield  {author} {\bibinfo {author} {\bibfnamefont {D.}~\bibnamefont {Sanchez-Manzano}}, \bibinfo {author} {\bibfnamefont {S.}~\bibnamefont {Mesoraca}}, \bibinfo {author} {\bibfnamefont {F.~A.}\ \bibnamefont {Cuellar}}, \bibinfo {author} {\bibfnamefont {M.}~\bibnamefont {Cabero}}, \bibinfo {author} {\bibfnamefont {V.}~\bibnamefont {Rouco}}, \bibinfo {author} {\bibfnamefont {G.}~\bibnamefont {Orfila}}, \bibinfo {author} {\bibfnamefont {X.}~\bibnamefont {Palermo}}, \bibinfo {author} {\bibfnamefont {A.}~\bibnamefont {Balan}}, \bibinfo {author} {\bibfnamefont {L.}~\bibnamefont {Marcano}}, \bibinfo {author} {\bibfnamefont {A.}~\bibnamefont {Sander}}, \bibinfo {author} {\bibfnamefont {M.}~\bibnamefont {Rocci}}, \bibinfo {author} {\bibfnamefont {J.}~\bibnamefont {Garcia-Barriocanal}}, \bibinfo {author} {\bibfnamefont {F.}~\bibnamefont {Gallego}}, \bibinfo {author} {\bibfnamefont {J.}~\bibnamefont {Tornos}}, \bibinfo {author} {\bibfnamefont {A.}~\bibnamefont {Rivera}}, \bibinfo {author} {\bibfnamefont
  {F.}~\bibnamefont {Mompean}}, \bibinfo {author} {\bibfnamefont {M.}~\bibnamefont {Garcia-Hernandez}}, \bibinfo {author} {\bibfnamefont {J.~M.}\ \bibnamefont {Gonzalez-Calbet}}, \bibinfo {author} {\bibfnamefont {C.}~\bibnamefont {Leon}}, \bibinfo {author} {\bibfnamefont {S.}~\bibnamefont {Valencia}}, \bibinfo {author} {\bibfnamefont {C.}~\bibnamefont {Feuillet-Palma}}, \bibinfo {author} {\bibfnamefont {N.}~\bibnamefont {Bergeal}}, \bibinfo {author} {\bibfnamefont {A.~I.}\ \bibnamefont {Buzdin}}, \bibinfo {author} {\bibfnamefont {J.}~\bibnamefont {Lesueur}}, \bibinfo {author} {\bibfnamefont {J.~E.}\ \bibnamefont {Villegas}},\ and\ \bibinfo {author} {\bibfnamefont {J.}~\bibnamefont {Santamaria}},\ }\bibfield  {title} {\bibinfo {title} {Extremely long-range, high-temperature josephson coupling across a half-metallic ferromagnet},\ }\href {https://doi.org/10.1038/s41563-021-01162-5} {\bibfield  {journal} {\bibinfo  {journal} {Nature Materials}\ }\textbf {\bibinfo {volume} {21}},\ \bibinfo {pages} {188} (\bibinfo
  {year} {2022})}\BibitemShut {NoStop}%
\bibitem [{\citenamefont {Todorov}\ \emph {et~al.}(2024)\citenamefont {Todorov}, \citenamefont {Dhillon},\ and\ \citenamefont {Mangeney}}]{TodorovDhillonMangeney2024}%
  \BibitemOpen
  \bibfield  {author} {\bibinfo {author} {\bibfnamefont {Y.}~\bibnamefont {Todorov}}, \bibinfo {author} {\bibfnamefont {S.}~\bibnamefont {Dhillon}},\ and\ \bibinfo {author} {\bibfnamefont {J.}~\bibnamefont {Mangeney}},\ }\bibfield  {title} {\bibinfo {title} {{THz quantum gap: exploring potential approaches for generating and detecting non-classical states of THz light}},\ }\href {https://doi.org/10.1515/nanoph-2023-0757} {\bibfield  {journal} {\bibinfo  {journal} {Nanophotonics}\ }\textbf {\bibinfo {volume} {13}},\ \bibinfo {pages} {1681} (\bibinfo {year} {2024})}\BibitemShut {NoStop}%
\bibitem [{\citenamefont {Frisk~Kockum}\ \emph {et~al.}(2019)\citenamefont {Frisk~Kockum}, \citenamefont {Miranowicz}, \citenamefont {De~Liberato}, \citenamefont {Savasta},\ and\ \citenamefont {Nori}}]{RevUSC_2_2019}%
  \BibitemOpen
  \bibfield  {author} {\bibinfo {author} {\bibfnamefont {A.}~\bibnamefont {Frisk~Kockum}}, \bibinfo {author} {\bibfnamefont {A.}~\bibnamefont {Miranowicz}}, \bibinfo {author} {\bibfnamefont {S.}~\bibnamefont {De~Liberato}}, \bibinfo {author} {\bibfnamefont {S.}~\bibnamefont {Savasta}},\ and\ \bibinfo {author} {\bibfnamefont {F.}~\bibnamefont {Nori}},\ }\bibfield  {title} {\bibinfo {title} {Ultrastrong coupling between light and matter},\ }\href {https://doi.org/10.1038/s42254-018-0006-2} {\bibfield  {journal} {\bibinfo  {journal} {Nature Reviews Physics}\ }\textbf {\bibinfo {volume} {1}},\ \bibinfo {pages} {19–40} (\bibinfo {year} {2019})}\BibitemShut {NoStop}%
\bibitem [{\citenamefont {Cohen-Tannoudji}\ \emph {et~al.}(2001)\citenamefont {Cohen-Tannoudji}, \citenamefont {Dupont-Roc},\ and\ \citenamefont {Grynberg}}]{Book_Cohen_Ph_At}%
  \BibitemOpen
  \bibfield  {author} {\bibinfo {author} {\bibfnamefont {C.}~\bibnamefont {Cohen-Tannoudji}}, \bibinfo {author} {\bibfnamefont {J.}~\bibnamefont {Dupont-Roc}},\ and\ \bibinfo {author} {\bibfnamefont {G.}~\bibnamefont {Grynberg}},\ }\href@noop {} {\emph {\bibinfo {title} {Photons et Atomes}}}\ (\bibinfo  {publisher} {EDP Sciences/CNRS Editions},\ \bibinfo {year} {2001})\BibitemShut {NoStop}%
\bibitem [{\citenamefont {Rousseaux}\ \emph {et~al.}(2023)\citenamefont {Rousseaux}, \citenamefont {Todorov}, \citenamefont {Vasanelli},\ and\ \citenamefont {Sirtori}}]{Rousseaux_PhysRevB125417_2023}%
  \BibitemOpen
  \bibfield  {author} {\bibinfo {author} {\bibfnamefont {B.}~\bibnamefont {Rousseaux}}, \bibinfo {author} {\bibfnamefont {Y.}~\bibnamefont {Todorov}}, \bibinfo {author} {\bibfnamefont {A.}~\bibnamefont {Vasanelli}},\ and\ \bibinfo {author} {\bibfnamefont {C.}~\bibnamefont {Sirtori}},\ }\bibfield  {title} {\bibinfo {title} {Phonon-mediated dark to bright plasmon conversion},\ }\href {https://doi.org/10.1103/PhysRevB.108.125417} {\bibfield  {journal} {\bibinfo  {journal} {Phys. Rev. B}\ }\textbf {\bibinfo {volume} {108}},\ \bibinfo {pages} {125417} (\bibinfo {year} {2023})}\BibitemShut {NoStop}%
\bibitem [{\citenamefont {Boventer}\ \emph {et~al.}(2023)\citenamefont {Boventer}, \citenamefont {Simensen}, \citenamefont {Brekke}, \citenamefont {Weides}, \citenamefont {Anane}, \citenamefont {Kl\"aui}, \citenamefont {Brataas},\ and\ \citenamefont {Lebrun}}]{Boventer_PhysRevApplied014071_2023}%
  \BibitemOpen
  \bibfield  {author} {\bibinfo {author} {\bibfnamefont {I.}~\bibnamefont {Boventer}}, \bibinfo {author} {\bibfnamefont {H.~T.}\ \bibnamefont {Simensen}}, \bibinfo {author} {\bibfnamefont {B.}~\bibnamefont {Brekke}}, \bibinfo {author} {\bibfnamefont {M.}~\bibnamefont {Weides}}, \bibinfo {author} {\bibfnamefont {A.}~\bibnamefont {Anane}}, \bibinfo {author} {\bibfnamefont {M.}~\bibnamefont {Kl\"aui}}, \bibinfo {author} {\bibfnamefont {A.}~\bibnamefont {Brataas}},\ and\ \bibinfo {author} {\bibfnamefont {R.}~\bibnamefont {Lebrun}},\ }\bibfield  {title} {\bibinfo {title} {Antiferromagnetic cavity magnon polaritons in collinear and canted phases of hematite},\ }\href {https://doi.org/10.1103/PhysRevApplied.19.014071} {\bibfield  {journal} {\bibinfo  {journal} {Phys. Rev. Appl.}\ }\textbf {\bibinfo {volume} {19}},\ \bibinfo {pages} {014071} (\bibinfo {year} {2023})}\BibitemShut {NoStop}%
\bibitem [{\citenamefont {Todorov}(2014)}]{Todorov_PhysRevB075115_2014}%
  \BibitemOpen
  \bibfield  {author} {\bibinfo {author} {\bibfnamefont {Y.}~\bibnamefont {Todorov}},\ }\bibfield  {title} {\bibinfo {title} {Dipolar quantum electrodynamics theory of the three-dimensional electron gas},\ }\href {https://doi.org/10.1103/PhysRevB.89.075115} {\bibfield  {journal} {\bibinfo  {journal} {Phys. Rev. B}\ }\textbf {\bibinfo {volume} {89}},\ \bibinfo {pages} {075115} (\bibinfo {year} {2014})}\BibitemShut {NoStop}%
\bibitem [{\citenamefont {Boehm}\ \emph {et~al.}(2003)\citenamefont {Boehm}, \citenamefont {Roessli}, \citenamefont {Schefer}, \citenamefont {Wills}, \citenamefont {Ouladdiaf}, \citenamefont {Leli\`evre-Berna}, \citenamefont {Staub},\ and\ \citenamefont {Petrakovskii}}]{Boehm03}%
  \BibitemOpen
  \bibfield  {author} {\bibinfo {author} {\bibfnamefont {M.}~\bibnamefont {Boehm}}, \bibinfo {author} {\bibfnamefont {B.}~\bibnamefont {Roessli}}, \bibinfo {author} {\bibfnamefont {J.}~\bibnamefont {Schefer}}, \bibinfo {author} {\bibfnamefont {A.~S.}\ \bibnamefont {Wills}}, \bibinfo {author} {\bibfnamefont {B.}~\bibnamefont {Ouladdiaf}}, \bibinfo {author} {\bibfnamefont {E.}~\bibnamefont {Leli\`evre-Berna}}, \bibinfo {author} {\bibfnamefont {U.}~\bibnamefont {Staub}},\ and\ \bibinfo {author} {\bibfnamefont {G.~A.}\ \bibnamefont {Petrakovskii}},\ }\bibfield  {title} {\bibinfo {title} {Complex magnetic ground state of ${\mathrm{cub}}_{2}{\mathrm{o}}_{4}$},\ }\href {https://doi.org/10.1103/PhysRevB.68.024405} {\bibfield  {journal} {\bibinfo  {journal} {Phys. Rev. B}\ }\textbf {\bibinfo {volume} {68}},\ \bibinfo {pages} {024405} (\bibinfo {year} {2003})}\BibitemShut {NoStop}%
\end{thebibliography}%

\end{document}